%% file: th-2004-022.tex
\documentclass[12pt]{article}

\usepackage{amsmath}
\usepackage{amssymb}
\usepackage{euscript}

\usepackage{epsfig}

\usepackage{cite}

\usepackage{alltt}

\def\oal{${\cal O}(\alpha)$}

\newcommand{\smartpap}{p\hskip-7pt\hbox{$^{^{(\!-\!)}}$}}
\newcommand{\beq}{\begin{equation}}
\newcommand{\eeq}{\end{equation}}
\newcommand{\winhac}{{\sf WINHAC}}
\newcommand{\horace}{{\sf HORACE}}

\begin{document}                     

\allowdisplaybreaks

\begin{titlepage}

\begin{flushright}
{\bf  CERN-PH-TH/2004-022
}

{\bf FNT/T 2004/02
}
\end{flushright}

\vspace{5mm}
\begin{center}
{\LARGE
Comparisons of the Monte Carlo programs \horace\ and \winhac\ 
for single-$W$-boson production at hadron colliders$^{\star}$
}
\end{center}

\vspace{1mm}
\begin{center}
{\large\bf  C.M.\ Carloni\ Calame$^{a,b}$, 
            S.\ Jadach$^{c,d}$, \\
            G.\ Montagna$^{b,a}$, 
            O.\ Nicrosini$^{a,b}$  
            {\it and} W.\ P\l{}aczek$^{e,d}$   
}

\vspace{4mm}
{\em
$^a$Istituto Nazionale di Fisica Nucleare, Sezione di Pavia\\
  Via A. Bassi 6, I-27100 Pavia, Italy.\\
$^b$Dipartimento di Fisica Nucleare e Teorica, Universit\`a di Pavia,\\ 
Via A. Bassi 6, I-27100 Pavia, Italy.\\
$^c$Institute of Nuclear Physics, Polish Academy of Sciences, \\
  ul. Radzikowskiego 152, 31-342 Cracow, Poland.\\
$^d$Department of Physics, CERN, Theory Division, 
    CH-1211 Geneva 23, Switzerland.\\ \vspace{1mm}
$^e$Institute of Physics, Jagellonian University,\\
   ul. Reymonta 4, 30-059 Cracow, Poland.\\ 
}

\end{center}

\vspace{10mm}
\begin{abstract}
We present the comparisons of two independent Monte Carlo event 
generators, \horace\ and \winhac, for single-$W$-boson production 
in hadronic collisions with multiphoton effects in leptonic $W$ decays.
These comparisons were performed first at the parton level with 
fixed quark-beams energy, and then at the hadron level for proton--proton
collisions at the LHC. In general, a good agreement between the two
programs has been found. Possible sources of differences in some
of the presented results are discussed. 
We also present and discuss the effects of including non-zero quark
masses for the main single-$W$-boson observables at the LHC.
\end{abstract}

\vspace{10mm}
\begin{center}
{\it To be submitted to Acta Physica Polonica B}
\end{center}

\vspace{8mm}
\footnoterule
\noindent
{\footnotesize
\begin{itemize}
\item[${}^{\star}$]
  Work partly supported by 
  the Polish Government grants
  KBN 2P03B00122 
  and KBN 5P03B09320, 
  the EC FP5 Centre of Excellence ``COPIRA'' under the contract 
  No.\ IST-2001-37259 
  and the EC FP5 contract HPRN-CT-2000-00149. 
\end{itemize}
}

\vspace{1mm}
\begin{flushleft}
{\bf CERN-PH-TH/2004-022\\
 Februrary 2004
}
\end{flushleft}

\end{titlepage}

\section{Introduction}
\label{sec:intro}

The Large Hadron Collider (LHC) currently under construction at CERN,
apart from its large discovery potential in high-energy physics, 
will also offer an opportunity for precision measurements of some 
important electroweak parameters, see e.g.~Ref.~\cite{LHCYR:2000}. 
Some of these measurements will be able
to surpass even the high precision achieved by electron--positron colliders,
such as LEP, see e.g.~Ref.~\cite{MC4LHC-Jadach:2003}. 
A particularly important role among these is played by the measurement of the 
$W$-boson mass and width. It will be performed during the first phase
of running the LHC -- with low luminosity. This improved measurement
of the basic $W$-boson parameters will narrow down the mass window for
the Higgs boson searches in the second phase of the LHC run and, of course,
it will provide a more stringent test of the Standard Model.
The main source of $W$-bosons at the LHC will be the process of 
single-$W$-boson production. 
To avoid the huge QCD background the LHC experiments will
look only at leptonic $W$-decay channels, excluding $\tau$'s, which decay
mainly hadronically. Huge rates of single-$W$ (and single-$Z$) 
events that will be produced at the LHC during its first phase of running can
also be exploited for measurements of parton distribution functions
(PDFs) in the proton and parton luminosities~\cite{Dittmar:1997}.  

In order to achieve the above experimental goals high-precision 
theoretical predictions are needed for the process of single-$W$-boson 
production with leptonic $W$ decays, provided in terms of a Monte Carlo (MC) 
event generator.
One of the goals of the CERN Workshop on Monte Carlo Tools for the LHC,
held in the summer of 2003 at CERN~\cite{MC4LHC:2003}, particularly its 
Electroweak Working Group, was to collect such event generators and 
to perform their basic comparative tests. The aim of these tests was to 
check both the technical and physical precision of the respective programs
and to provide guidelines for their further improvements.
None of the currenlty existing MC event generators includes all the ingredients
that are necessary for the real experimental data analysis of the single-$W$
events. They can, generally, be split into two classes: the programs that deal
with the QCD effects and those dedicated to a precision description of
QED/EW corrections~\cite{MC4LHC-Jadach:2003}.

Here we consider two MC event generators: \horace~\cite{HORACE:2003} 
and \winhac~\cite{WINHAC:MC}, which in their current versions
include higher-order QED corrections in leptonic $W$-boson decays. 
Control over the QED correction in $W$ decays is of crucial importance
for the precision determination of the $W$ mass and width at hadron
colliders, see e.g.~\cite{Baur:1998}. \winhac\ and \horace\ are the
first MC event generators to go beyond \oal\ in accounting for these
corrections.

We performed numerical comparisons of the above programs in two steps.
In the first step, we compared their predictions at the pure parton level,
for fixed initial quark energies. The processes considered in this step
do not correspond directly to the ones studied in real-life experiments; 
however, such tests are useful for technical reasons. 
The QED corrections in leptonic $W$-boson decays 
are actually included in the parton-level processes;
testing them at this level thus allows a direct check of their implementation,
without any influence of additional effects (PDFs, QCD, etc.). 
One should, however, keep in mind that any differences seen at this level 
have a rather technical meaning and cannot be simply translated into 
realistic proton--proton collisions.

In the second step, the two programs were cross-checked at the hadron level,
i.e. for proton--proton collisions at the LHC energies, where the basic
parton-level hard processes were convoluted with the PDFs of the two colliding
protons. In these comparisons we considered four single-$W$ quantities:
the $W$-boson transverse mass, the charged lepton transverse momentum,
the $W$-boson pseudorapidity%
   \footnote{The $W$-boson pseudorapidity is not measured in the real
             experiment since neutrino detection is lacking, however
             it can be useful for some generator-level studies.}
and the charged lepton pseudorapidity.
The first two are sensitive to the $W$-boson mass and width, the
other two to the PDFs and the parton luminosities. 
In addition, we compared two photon observables: the hardest-photon transverse
momentum and its pseudorapidity. 

The paper is organized as follows. In Section 2 we briefly describe the
two programs. In Sections 3 and 4 we present and discuss
the comparisons at the parton level and at the hadron level, respectively.
Section 5 is devoted to quark-mass effects on the main $W$-boson
observables. Finally, Section 6 contains a summary and outlook of this work.

\section{Programs}
\label{sec:progs}

\subsection{HORACE}
\label{ssec:horace}
\horace~\cite{HORACE:2003} is a Monte Carlo event generator for the
Drell--Yan-like processes  $p\smartpap \to W \to l \nu_l$
and $p\smartpap \to \gamma,Z \to l^+ l^-$, with $l=e,\mu$.
In \horace \ the corrections due to (real plus virtual)
multiphoton radiation are computed using the well-known 
QED structure-function approach. The
corrections are calculated by solving numerically the DGLAP evolution equation 
for the QED structure function by means of the parton shower algorithm
described in detail in Refs.~\cite{PS-PV:2000,PS-C:2001}. 

Only the corrections relative to the subprocesses $W \to l \nu_l$ and 
$Z \to l^+ l^-$  are at present included in \horace, because it is known that 
quark-mass singularities,
originating from initial-state photon radiation, can be reabsorbed into a 
redefinition of the PDFs, in analogy to gluon emission in QCD.
\horace \ can calculate photonic corrections to all orders and at \oal, 
to disentangle the effect of higher-order contributions and to compare it with
that of the available \oal \ programs. Technically speaking, the energy of the 
emitted photons is extracted from the Altarelli--Parisi $e \to e + \gamma$ 
splitting function, while the
generation of the photons' angles is performed according to the factorized 
part of the radiation matrix element, which for the $W \to l \nu \gamma$ 
decay reads as~\cite{Berends:1985}
$(p/p \cdot k - Q/Q \cdot k)^2$, where $p$ is the four-momentum of the 
final-state lepton $l$, 
$k$ is the photon four-momentum, and $Q$ is the four-momentum (virtuality) 
of the $W$ boson.

Complete \oal \ electroweak corrections are not 
included in the present version of the generator, but work is in progress
in order to make them available in a future release. To provide predictions 
at the hadron level, the parton-level subprocesses for $W$ and $Z$ production
in quark--antiquark annihilation are convoluted with collinear PDFs from the 
PDFLIB package~\cite{PDFLIB:2000}.


\subsection{WINHAC}
\label{ssec:winhac}

\winhac~\cite{WINHAC:MC} is a Monte Carlo event generator
for single-$W$-boson production in hadronic collisions (proton--proton
and proton--antiproton) with leptonic $W$ decays. Its main feature is
the Yennie--Frautschi--Suura (YFS) \oal\ QED exponentiation in leptonic 
$W$-boson decays, described in detail in Ref.~\cite{WINHAC:2003}. 
\winhac\ generates multiphoton radiation in $W$ decays, which is exact: 
(1) to all orders in the infrared (IR) limit and 
(2) up to \oal\ for non-IR contributions.
The non-IR photonic contributions beyond \oal\ are included in an approximate
way: some parts of leading-log as well as subleading terms are taken into
account through the YFS exponentiation.
The recent version of the program includes also the \oal\ electroweak (EW)
corrections to the leptonic $W$ decays~\cite{Bardin:2003}, 
implemented within the YFS exponentiation scheme. The implementation of
the full \oal\ EW corrections to the charged-current Drell--Yan process 
is under way.

The parton-level hard process of single-$W$ production in quark--antiquark
collisions is convoluted with the standard proton PDFs from the PDFLIB
package~\cite{PDFLIB:2000} -- to give a hadron--level process 
appropriate for proton--proton or proton--antiproton collisions. 
The Bjorken $x$'s and $Q^2$ of the quark--antiquark pair 
are generated with the help of the self-adaptive MC sampler 
Foam~\cite{Foam:2003}. QCD effects are included in the current version
only trough scaling-violation of collinear PDFs (no quark $p_T$, 
no parton showers, no hadronization).

\section{Parton-level comparisons}
\label{sec:parlev}
For the parton-level comparisons we considered the following processes:
\begin{equation}
d + \bar{u} \longrightarrow W^-  \longrightarrow l^- + \bar{\nu_l}, 
\hspace{4mm} l=e,\,\mu,
\label{eq:parpro}
\end{equation}
with the $+z$ axis pointing in the incoming $d$-quark direction.
Our MC calculations were done in the $G_{\mu}$ scheme and the fixed-width
scheme, for the following input parameters:
\begin{equation}
\begin{aligned}
\,& m_d = 3\times 10^{-3}\,{\rm GeV},\:\:\:\:\:\: m_u = 6\times 10^{-3},
  \:\:\:\:\:\: V_{ud} = 1, \:\:\:\:\:\: m_{\nu_l} = 0,\\
\,& m_e = 0.511\times 10^{-3}\,{\rm GeV},\:\:\:\:\:\:
  m_{\mu} = 0.10565836\,{\rm GeV},\:\:\:\:\:\:
  m_{\tau} = 1.77703\,{\rm GeV}, \\
\,& M_W = 80.423\,{\rm GeV},\:\:\:\:\:\:  M_Z = 91.1882\,{\rm GeV}\\
\,& s_W^2 = 1 -\frac{M_W^2}{M_Z^2}, \:\:\:\:\:\: 
\Gamma_W = \frac{3G_{\mu}M_W^3}{2\sqrt{2}\pi}\left(1 
           + \frac{2\alpha_s}{3\pi}\right), \\
\,& \alpha^{-1} = 137.03599976,\:\:\:\:\:\:
G_{\mu} = 1.16639 \times 10^{-5}\,{\rm GeV}^{-2}, \:\:\:\:\:\:
\alpha_s = 0.1185, \\
\,& E_{\rm CM} = \sqrt{s} = M_W. \\
\end{aligned}
\label{eq:parinput}
\end{equation}
The results have been obtained for two kinds of event selection: 
\begin{itemize}
\item
  {\bf BARE} -- where the corresponding observables
  were obtained from bare-lepton four-momenta and no cuts were applied.
\item 
  {\bf CALO} -- where the photon four-momenta were combined
  with the charged-lepton four-momenta if the opening angle between their
  directions $\angle(\vec{q}_l,\vec{k})\le 5^{\circ}$; such photons were
  discarded; no extra cuts were applied. 
\end{itemize}

We considered only QED-like corrections to leptonic $W$-boson decays.
Our computations were done at three levels: (1) Born, (2) \oal\ (as implemented
in the respective programs), and (3) with higher-order corrections (the
so-called ``Best'' predictions). 

\begin{table}[!ht]
\centering
\begin{tabular}{||c|c|c|c||}
\hline\hline
\raisebox{-1.5ex}[0cm][0cm]{Program} & 
\multicolumn{3}{|c||}{$\sigma^{\rm tot}\,$[nb]} \\
\cline{2-4}
     & Born           & \oal         & Best \\
\hline\hline
  \multicolumn{4}{||c||}{Electrons} \\
\cline{1-4}
\horace   & $8.88722\,(00)$ & $8.88721\,(00)$ & $8.88721\,(0)$ \\
\winhac   & $8.88715\,(20)$ & $8.88552\,(12)$ & $8.88401\,(5)$ \\
\hline
$\delta=$ {\sf (W $-$ H)/W}
  & $-0.8\,(2.3)\times 10^{-5}$
  & $-1.9\,(0.1)\times 10^{-4}$ 
  & $-3.60\,(0.06)\times 10^{-4}$ \\
\hline\hline
  \multicolumn{4}{||c||}{Muons} \\
\cline{1-4}
\horace   & $8.88722\,(00)$ & $8.88632\,(1)$ & $8.88632\,(1)$ \\
\winhac   & $8.88720\,(13)$ & $8.88533\,(6)$ & $8.88440\,(5)$ \\
\hline
$\delta=$ {\sf (W $-$ H)/W}
  & $-0.2\,(1.4)\times 10^{-5}$
  & $-1.11\,(0.07)\times 10^{-4}$ 
  & $-2.16\,(0.06)\times 10^{-4}$ \\
\hline\hline
\end{tabular}
\caption{\small\sf
  The parton-level results for the Born, the \oal\ and the ``Best''-level total
  cross section in nanobarns [nb] from \horace\ and \winhac.
  The numbers in parentheses are statistical errors 
  for the last digits.
}
\label{tab:xtot-par}
\end{table}
In Table~\ref{tab:xtot-par} we present the results for the total
cross section from \horace\ and \winhac. As can be seen, the 
agreement between the two programs for the total integrated
cross section without any cut is excellent, the 
relative differences being at the level of $10^{-4}$ both at the
Born level and in the presence of QED corrections.
The remaining differences of $\sim 2\times 10^{-4}$ for the \oal\ and ``Best''
predictions can be explained by the different treatment of QED corrections
in the structure-functions (SF) approach and the YFS exclusive exponentiation.
In the SF calculations the net effect of the QED correction for the total
cross section is zero (except for some small lepton-mass effects in the 
phase-space integration). In the YFS exponentiation, on the other hand,
the QED corrections to the total cross section are non-zero due to subleading
terms, which are at the level of $\sim 2\times 10^{-4}$, as was shown in
Ref.~\cite{WINHAC:2003}.  

In the following we present the comparisons of various distributions.
As a first step, we compared the Born-level distributions of lepton energy 
and polar angle for both the electron and muon channels, obtaining
a very good agreement (within statistical errors corresponding to the
same event samples, as presented in Table 1).
The lepton energy and lepton angle distributions 
for \oal\ and ``Best'' predictions of the
two programs, and when considering the BARE and CALO event selections, 
are shown in Figs.~\ref{fig:Par_el_El}--\ref{fig:Par_mu_costhl}. 
For such distributions, the agreement between the two programs 
can be considered generally good. Actually, the differences are at the level of
$1\%$ in those regions of the differential distributions, yielding
the dominant contribution to the integrated cross section, 
while they can reach the $10\%$ (or more) level only 
in the kinematical regions that are dynamically and/or kinematically
suppressed. 
In Figs.~\ref{fig:Par_el_Eg}--\ref{fig:Par_mu_costhg} we present
distributions of photonic observables: the hardest photon energy,
the total photon energy (i.e.\ the sum of all radiative-photon energies),
and the hardest photon polar angle. As for the leptonic distributions,
the agreement is quite satisfactory since, whenever ``large'' differences
are present, they occur in those regions that give only a small
contribution to the integrated cross section. In particular, it can
be seen from Figs.~\ref{fig:Par_el_Eg}--\ref{fig:Par_mu_Egtot} that
the agreeement is considerably better for the CALO event selection
than for the BARE one, as expected as a consequence of 
the (partial) cancellation of leading-logarithmic corrections
in the more inclusive CALO case. 

As can be seen in Figs.~\ref{fig:Par_el_El}--\ref{fig:Par_mu_Egtot},
the differences between \horace\ and \winhac\ for their ``Best'' predictions
are almost the same as the ones at \oal. This indicates that the main
effect comes from different treatment of the \oal\ subleading terms in the
two programs. The differences become large in the regions of phase space
where these subleading terms start to dominate. In these regions, however,
the cross section is very small.

%

\section{Hadron-level comparisons}
\label{sec:hadlev}

For the hadron-level comparisons we considered the following processes:
\begin{equation}
\begin{aligned}
pp & \longrightarrow W^- + X  \longrightarrow l^- + \bar{\nu_l} + X, \\
pp & \longrightarrow W^+ + X  \longrightarrow l^+ + \nu_l + X, 
\end{aligned}
\label{eq:hadpro}
\end{equation}
where $l=e,\,\mu$. In this case, the parton-level processes of $W$ production
and decay were convoluted with the standard PDFs.
For our tests we used the MRS(G) parametrization as provided
by the PDFLIB package~\cite{PDFLIB:2000} 
({\tt Ntype = 1, Ngroup = 3, Nset = 41}).
As in the parton-level case, our MC calculations were done in the $G_{\mu}$ 
scheme and the fixed-width scheme, for the following input parameters:
\begin{equation}
\begin{aligned}
\,& m_e = 0.511\times 10^{-3}\,{\rm GeV},\:\:\:\:\:\:
  m_{\mu} = 0.10565836\,{\rm GeV}, \:\:\:\:\:\: m_{\nu_e} = m_{\nu_{\mu}}=0 \\
\,& m_u = m_d = m_s = m_c = m_b = 0, \:\:\:\: m_t = 174.3\,{\rm GeV},
\:\:\:\:\:\: m_H = 150\,{\rm GeV},\\
\,& V_{ud} =  0.97483, \:\:\:\:\:\:\:\: V_{us} =  0.22290,  \:\:\:\:\:\:\:\:
     V_{ub} = 0.00360,\\
\,& V_{cd} = -0.22286, \:\:\:\:\: V_{cs} =  0.97398,  \:\:\:\:\:\:\:\: 
    V_{cb} = 0.04120, \\
\,& V_{td} =  0.00568, \:\:\:\:\:\:\:\:\; V_{ts} = -0.04097,  \:\:\:\:\:\, 
    V_{tb} = 0.99914, \\
\,& M_W = 80.423\,{\rm GeV},\:\:\:\:\:\:  M_Z = 91.1882\,{\rm GeV}\\
\,& s_W^2 = 1 -\frac{M_W^2}{M_Z^2}, \:\:\:\:\:\: 
\Gamma_W = \frac{3G_{\mu}M_W^3}{2\sqrt{2}\pi}\left(1 
           + \frac{2\alpha_s}{3\pi}\right), \\
\,& \alpha^{-1} = 137.03599976,\:\:\:\:\:\:
G_{\mu} = 1.16639 \times 10^{-5}\,{\rm GeV}^{-2}, \:\:\:\:\:\:
\alpha_s = 0.1185, \\
\,& E_{\rm CM} = \sqrt{s} = 14\,{\rm GeV}. \\
\end{aligned}
\label{eq:hadinput}
\end{equation}
We used the following event selection criteria: 
\begin{itemize}
\item
  the charged lepton transverse momentum:  $p_T^l > 25\, {\rm GeV}$;
\item
  the charged lepton pseudorapidity:      $|\eta_l| < 2.4$;
\item
  the missing transverse energy:         $ E_T^{\rm miss} > 25\,{\rm GeV}$
    (we used $E_T^{\rm miss} = p_T^{\nu}$);
\item
  the size of an electron cluster (for electron--photon recombination): 
    $\Delta\eta_e \times \Delta\phi_e = 0.075\times 0.175\,{\rm rad}$,
    where $\eta_e$ and $\phi_e$ are the electron pseudorapidity and azimuthal 
    angle;
\item
 no photon recombination with muons.
\end{itemize}
For our main tests we chose four basic single-$W$ observables:
\begin{enumerate}
\item
  $W$-boson transverse mass: $m_T^W$;
\item
  charged lepton transverse momentum: $p_T^l$;
\item 
  $W$-boson rapididy: $y_W$;
\item 
   charged lepton pseudorapidity: $\eta_l$.
\end{enumerate}
The first two are important for the $W$ mass and width measurement, 
the other two for the PDFs and parton-luminosity measurements.

As additional tests, we performed the comparisons of photonic distributions:
\begin{enumerate}
\item
  hardest photon transverse momentum: $p_T^{\gamma}$;
\item 
  hardest photon pseudorapidity: $\eta_{\gamma}$,
\end{enumerate}
for which we used additional cuts:
\begin{itemize}
\item
  photon transverse momentum:  $p_T^ {\gamma}> 25\,{\rm GeV}$;
\item
  photon pseudorapidity:      $|\eta_{\gamma}| < 2.4$.
\end{itemize}

\begin{table}[!ht]
\centering
\begin{tabular}{||c|c|c|c||}
\hline\hline
\raisebox{-1.5ex}[0cm][0cm]{Program} & 
\multicolumn{3}{|c||}{$\sigma^{\rm tot}\,$[nb]: WITHOUT CUTS} \\
\cline{2-4}
     & Born           & \oal         & Best \\
\hline\hline
  \multicolumn{4}{||c||}{$W^- \longrightarrow e^-\bar{\nu}_e$} \\
\cline{1-4}
\horace   & $7.73310\,(39)$ & $7.73314\,(41)$ & $7.73249\,(43)$ \\
\winhac   & $7.73315\,(11)$ & $7.73171\,(07)$ & $7.73039\,(03)$ \\
\hline
$\delta=$ {\sf (W $-$ H)/W}
  & $ 0.6\,(5.2)\times 10^{-5} $
  & $-1.8\,(0.5)\times 10^{-4} $ 
  & $-2.7\,(0.5)\times 10^{-4} $ \\
\hline\hline
  \multicolumn{4}{||c||}{$W^- \longrightarrow \mu^-\bar{\nu}_{\mu}$} \\
\cline{1-4}
\horace   & $7.73317\,(39)$ & $7.73322\,(39)$ & $7.73285\,(41)$ \\
\winhac   & $7.73317\,(07)$ & $7.73162\,(04)$ & $7.73075\,(03)$ \\
\hline
$\delta=$ {\sf (W $-$ H)/W}
  & $ 0.0\,(5.1)\times 10^{-5} $
  & $-2.1\,(0.5)\times 10^{-4} $ 
  & $-2.7\,(0.5)\times 10^{-3} $ \\
\hline\hline
  \multicolumn{4}{||c||}{$W^+ \longrightarrow e^+\nu_e$} \\
\cline{1-4}
\horace   & $10.93760\,(56)$ & $10.93804\,(59)$ & $10.93679\,(61)$ \\
\winhac   & $10.93684\,(17)$ & $10.93535\,(12)$ & $10.93322\,(04)$ \\
\hline
$\delta=$ {\sf (W $-$ H)/W}
  & $-6.9\,(5.4)\times 10^{-5}$
  & $-2.5\,(0.6)\times 10^{-4}$ 
  & $-3.3\,(0.6)\times 10^{-4}$ \\
\hline\hline
  \multicolumn{4}{||c||}{$W^+ \longrightarrow \mu^+\nu_{\mu}$} \\
\cline{1-4}
\horace   & $10.93757\,(56)$ & $10.93856\,(57)$ & $10.93683\,(57)$ \\
\winhac   & $10.93714\,(11)$ & $10.93507\,(06)$ & $10.93381\,(04)$ \\
\hline
$\delta=$ {\sf (W $-$ H)/W}
  & $-3.9\,(5.2)\times 10^{-5} $
  & $-3.2\,(0.5)\times 10^{-4}$ 
  & $-2.8\,(0.5)\times 10^{-4}$ \\
\hline\hline
\end{tabular}
\caption{\small\sf
  The hadron-level results for the Born, the \oal\ and the ``Best''-level total
  cross section (in nb) from \horace\ and \winhac\ without cuts.
  The numbers in parentheses are statistical errors 
  for the last digits.
}
\label{tab:xtot-had}
\end{table}

\begin{table}[!ht]
\centering
\begin{tabular}{||c|c|c|c||}
\hline\hline
\raisebox{-1.5ex}[0cm][0cm]{Program} & 
\multicolumn{3}{|c||}{$\sigma^{\rm tot}\,$[nb]:  WITH CUTS} \\
\cline{2-4}
     & Born           & \oal         & Best \\
\hline\hline
  \multicolumn{4}{||c||}{$W^- \longrightarrow e^-\bar{\nu}_e$} \\
\cline{1-4}
\horace   & $3.23633\,(12)$ & $3.18707\,(13)$ & $3.18696\,(13)$ \\
\winhac   & $3.23629\,(09)$ & $3.18779\,(07)$ & $3.18765\,(06)$ \\
\hline
$\delta=$ {\sf (W $-$ H)/W}
  & $-1.2\,(4.6)\times 10^{-5} $
  & $ 2.3\,(0.5)\times 10^{-4} $ 
  & $ 2.2\,(0.5)\times 10^{-4} $ \\
\hline\hline
  \multicolumn{4}{||c||}{$W^- \longrightarrow \mu^-\bar{\nu}_{\mu}$} \\
\cline{1-4}
\horace   & $3.23632\,(12)$ & $3.15990\,(12)$ & $3.16013\,(13)$ \\
\winhac   & $3.23630\,(07)$ & $3.16418\,(06)$ & $3.16409\,(05)$ \\
\hline
$\delta=$ {\sf (W $-$ H)/W}
  & $-0.6\,(4.3)\times 10^{-5} $
  & $ 1.35\,(0.05)\times 10^{-3} $ 
  & $ 1.25\,(0.05)\times 10^{-3} $ \\
\hline\hline
  \multicolumn{4}{||c||}{$W^+ \longrightarrow e^+\nu_e$} \\
\cline{1-4}
\horace   & $4.39341\,(16)$ & $4.32186\,(17)$ & $4.32187\,(18)$ \\
\winhac   & $4.39328\,(13)$ & $4.32286\,(10)$ & $4.32273\,(08)$ \\
\hline
$\delta=$ {\sf (W $-$ H)/W}
  & $-3.0\,(4.7)\times 10^{-5}$
  & $ 2.3\,(0.5)\times 10^{-4}$ 
  & $ 2.0\,(0.5)\times 10^{-4}$ \\
\hline\hline
  \multicolumn{4}{||c||}{$W^+ \longrightarrow \mu^+\nu_{\mu}$} \\
\cline{1-4}
\horace   & $4.39340\,(16)$ & $4.28255\,(16)$ & $4.28326\,(16)$ \\
\winhac   & $4.39336\,(10)$ & $4.28837\,(08)$ & $4.28848\,(08)$ \\
\hline
$\delta=$ {\sf (W $-$ H)/W}
  & $-0.9\,(4.3)\times 10^{-5} $
  & $ 1.36\,(0.05)\times 10^{-3}$ 
  & $ 1.22\,(0.05)\times 10^{-3}$ \\
\hline\hline
\end{tabular}
\caption{\small\sf
  The hadron-level results for the Born, the \oal\ and the ``Best''-level total
  cross section (in nb) from \horace\ and \winhac\ with cuts.
  The numbers in parentheses are statistical errors 
  for the last digits.
}
\label{tab:xtot-had_cuts}
\end{table}
In Tables~\ref{tab:xtot-had} and \ref{tab:xtot-had_cuts} we present 
the results for the total cross section from \horace\ and \winhac\ 
without cuts and with cuts (as described above), respectively. 
%
Except for the muon channels in the presence of the cuts, where the
differences between the two programs are at the level of $\sim 10^{-3}$,
the agreement between the two programs is consistent with the one at the
parton level. The differences in the muon channels can be explained
by the muon mass effects enhanced by the cuts. In \winhac\ the non-zero
muon masses are taken into account, while in \horace\ the muon masses are
neglected.
It can also be noticed that the differences
observed at the level of \oal\ cross sections still remain 
in the presence of higher-order corrections (the ``Best'' predictions).
This clearly indicates a difference in the treatment of 
\oal\ subleading corrections, which is expected on the grounds
of the different ingredients and formulation of the two generators,
and an almost complete agreement for the size of higher-order
contributions. 

The comparisons of the distributions are shown in 
Figs.~\ref{fig:Had_Wm_MTW}--\ref{fig:Had_Wp_etag}. The
results for both the $W^+$ and $W^-$ decays are shown. For the
transverse mass and lepton transverse momentum distributions,
as well as for the $W$ rapidity and lepton pseudorapidity 
distributions, the differences are confined within $0.2\%$, 
indicating a very good agreement between the two programs
for the observables relevant to the measurement of
the $W$ mass and width and for the parton-luminosity
determination. For the electron channels the agreement is generally
within the statistical errors, while for the muon channels there
are systematic differences at the level $0.1$--$0.2\%$. They can
be explained by the neglect of muon masses in \horace, as in the case
of the total cross section. 
For the photonic distributions, the
differences can reach some per-cent level, 
which can again be ascribed to 
the different treatment of \oal\ subleading terms.

In Figs.~\ref{fig:Had_Wm_MTW_Cor}--\ref{fig:Had_Wm_etag_Cor} we
show the size of the \oal\ and higher-order QED corrections for the
above distributions. As can be seen, the predictions of 
\oal\ and higher-order QED corrections from the 
two programs are in good agreement, in both shape and size. It can
in particular be observed that the contribution of QED corrections 
is almost flat for the $W$-boson rapidity and lepton pseudorapidity,
as well as for the exclusive photon observables, whereas it is a 
varying function for the $W$-boson transverse mass spectrum and 
the lepton transverse momentum distribution. 
In the two latter cases, the \oal\ corrections 
around the distribution peaks amount to about $5\%$ and about 
$10\%$ for the electron and muon channel, respectively, 
while higher-order effects vary from $0.2\%$ to $0.5\%$. 

\section{Quark-mass effects}
\label{sec:qmass}

In this section we present the effects of using non-zero quark masses
in the hadron-level $W$-production process for various observables.
In our test we used the following quark masses:
\begin{equation}
\begin{aligned}
\,& m_u =  0.003\,{\rm GeV}, \:\:\:\: m_d = 0.00675\,{\rm GeV}, \:\:\:\:
    m_s = 0.1175\,{\rm GeV}\\
\,& m_c = 1.2\,{\rm GeV}, \:\:\:\:\:\:\:\: m_b = 4.25\,{\rm GeV}.
\end{aligned}
\label{eq:qmass}
\end{equation}
All other input parameters were the same as in the previous section.

In the case of non-zero quark masses, there is an ambiguity in defining
the quark four-momentum $q$ in terms of proton momentum $p$ and
the Bjorken $x$-variable. Some of the possible definitions are:
\begin{enumerate}
\item the ``energy-like scheme'' : $ q^0 = x p^0 $;
\item the ``momentum-like scheme'': $ q^3 = x p^3 $;
\item the ``light-cone-like scheme'': $ q^+ = x p^+$, 
          where $a^+ =(a^0 + a^3)/\sqrt{2}$. 
\end{enumerate}
All the above definitions are equivalent for zero quark masses.
 
In the following figures we compare the distributions presented 
in the previous section for zero quark masses with the corresponding ones 
obtained using non-zero quark masses. 
The Figures~\ref{fig:xen_Wm_MTW}--\ref{fig:xen_Wm_etal} show the size
of the quark-mass effect for the ``energy-like scheme'', 
Figs.~\ref{fig:xpz_Wm_MTW}--\ref{fig:xpz_Wm_etal} for the 
``momentum-like scheme'', 
and Figs.~\ref{fig:xlc_Wm_MTW}--\ref{fig:xlc_Wm_etal} for the 
``light-cone-like scheme''. 
It is worth noticing that the size and also the sign 
of the quark-mass effects significantly depend on the scheme adopted.
For example, in the ``energy-like scheme'' the quark-mass 
effects are of the order of $+2\%$ around the peaks of
the transverse mass and lepton transverse momentum distributions, while
in the ``momentum-like'' scheme the corrections are of the same
order but of opposite sign, amounting to about $-3\%$. These considerations
about the different sign of the effects predicted by the above two schemes
do apply to the $W$ rapidity and lepton pseudorapidity, even if 
for such distributions the effect is confined at the few per mille level.
The smallest effects due to non-zero quark masses is observed 
for all the considered distributions in the ``light-cone-like'' scheme.
An appropriate inclusion of non-zero quark mass effects should therefore
be carefully considered in view of future improved measurements 
of the $W$-boson mass and width at hadron colliders.

\section{Summary and outlook}
\label{sec:summary}

We presented a number of tuned comparisons between the predictions of
two recently developed event generators, \horace\ and \winhac\, 
for single-$W$-boson production at hadron colliders. This is important
in view of future precision determinations of the $W$-boson mass and
width, as well as for PDFs and parton-luminosity measurements.

Both generators include \oal\ and higher-order QED corrections
to leptonic $W$ decays, according to different and independent
formulations. The comparisons were performed both at the parton level
with fixed quarks-beam energy and at the hadron level for proton--proton
collisions at the LHC. To understand the source of possible discrepancies,
the comparisons were carried out at three different levels of theoretical
precision:  Born, \oal\ and with higher-order effects. Both integrated 
cross sections and differential distributions were examined.
%
The agreement between the two programs for the main 
single-$W$-boson observables is satisfactory. 
The higher-order QED corrections, although based on two quite different
approaches, numerically agree very well. 
These comparisons indicate that both \horace\ and \winhac\
describe the QED effects in leptonic $W$-boson decays with the precision
that is sufficient for the $W$-boson mass and width determination 
as well as the PDFs and parton luminosities measurements at the LHC.
Of course, neither of the programs includes all the effects that are 
necessary for the full experimental data analysis; however, they are
good starting points for further developments.

The next step would be to see how various QED effects and differences
between the programs translate into fitted values of the $W$ mass and width
and the actual determination of the PDFs and parton luminosities. 
%
%
A first evaluation of the impact of higher-order QED corrections to the
$W$ mass extraction, from a fit to the transverse-mass distribution,
has been performed in Ref.~\cite{HORACE:2003}. 
A more realistic analysis should, however, be made in close connection 
with the actual experimental procedures.

Further tests should also include the complete 
\oal\ EW corrections for single-$W$-boson production, and, last but not least, 
the QCD effects, which are crucial for any precise measurement at the LHC.

In last section, we show how the inclusion of non-zero quark 
masses may affect the main single-$W$-boson observables at the LHC. 
Both the size and shape of these effects depend on the scheme adopted for
dealing with massive-quarks kinematics. Particularly large effects are
seen in the distributions of the $W$-boson transverse mass and the charge 
lepton transverse momentum. Therefore, the proper treatment of quark masses 
in theoretical descriptions of the respective processes can be of
importance for the precision measurement of the $W$ mass and width at
the LHC.   

We plan to carry out similar studies for proton--antiproton collisions at 
the Tevatron.


\vspace{5mm}
\noindent
{\large\bf Acknowledgements}
\vspace{3mm}

\noindent
Two of us (S.J. and W.P.) acknowledge the kind support of the CERN 
TH Unit, Physics Department.

\include{pardif}
%

\include{haddif}

%
\include{hadcor}

\include{hadxen}
\include{hadxpz}
\include{hadxlc}

\end{document}

%% file: pardif.tex
\begin{figure}[!ht]
\setlength{\unitlength}{1mm}
\begin{picture}(160,75)
\put( -3,0){\makebox(0,0)[lb]{
\epsfig{file=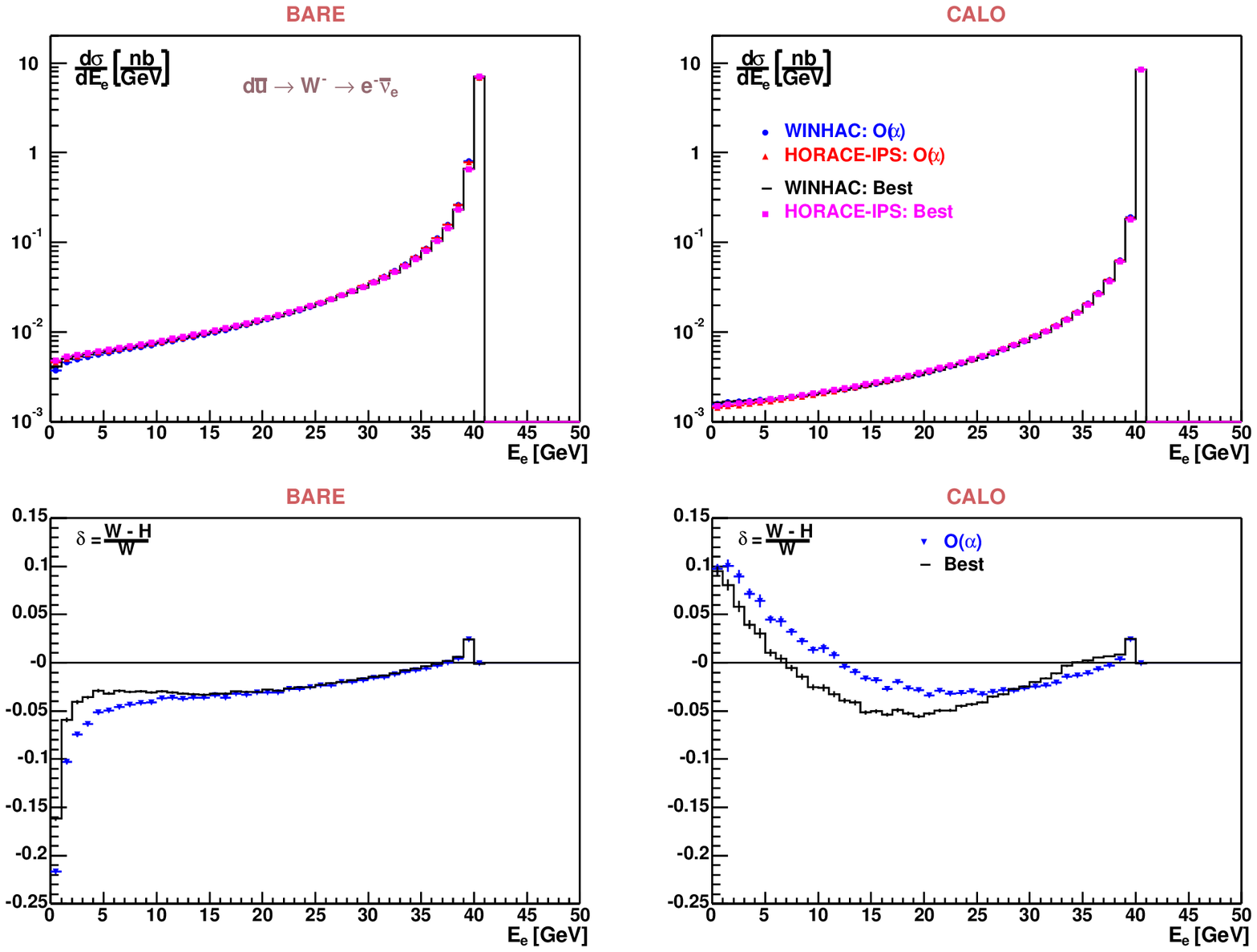,width=168mm,height=88mm}
}}
\end{picture}
\caption{\small\sf
Distributions of the electron energy at parton level
for \oal\ and ``Best'' predictions from 
\horace\ and \winhac, as well as their differences. The results are shown
for BARE (left) and CALO (right) event selections.
}
\label{fig:Par_el_El}
\end{figure}
%
\begin{figure}[!ht]
\setlength{\unitlength}{1mm}
\begin{picture}(160,88)
\put( -3,0){\makebox(0,0)[lb]{
\epsfig{file=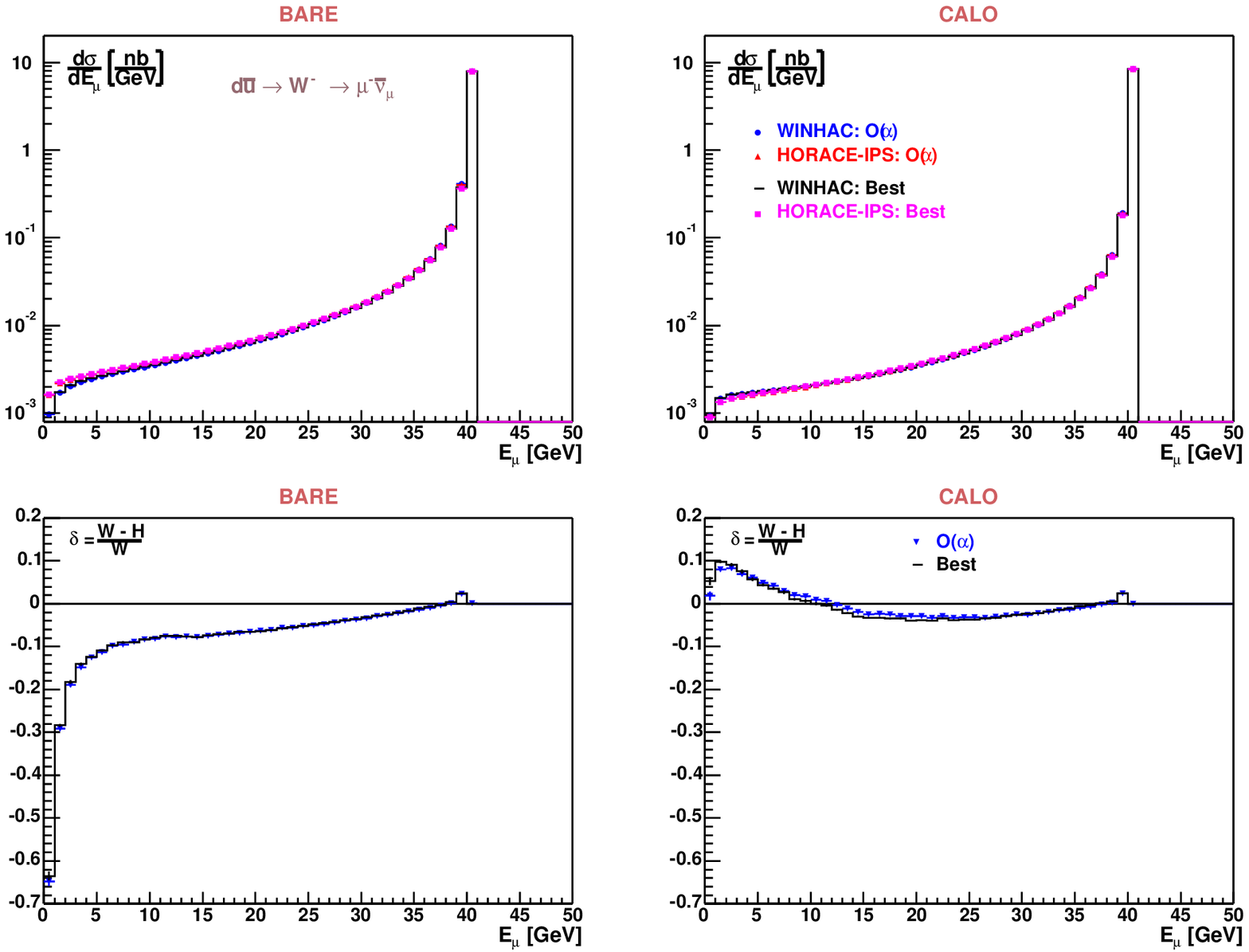,width=168mm,height=88mm}
}}
\end{picture}
\caption{\small\sf
Distributions of the muon energy at parton level 
for \oal\ and ``Best'' predictions from 
\horace\ and \winhac, as well as their differences. The results are shown
for BARE (left) and CALO (right) event selections.
}
\label{fig:Par_mu_El}
\end{figure}
%
\begin{figure}[!ht]
\setlength{\unitlength}{1mm}
\begin{picture}(160,88)
\put( -3,0){\makebox(0,0)[lb]{
\epsfig{file=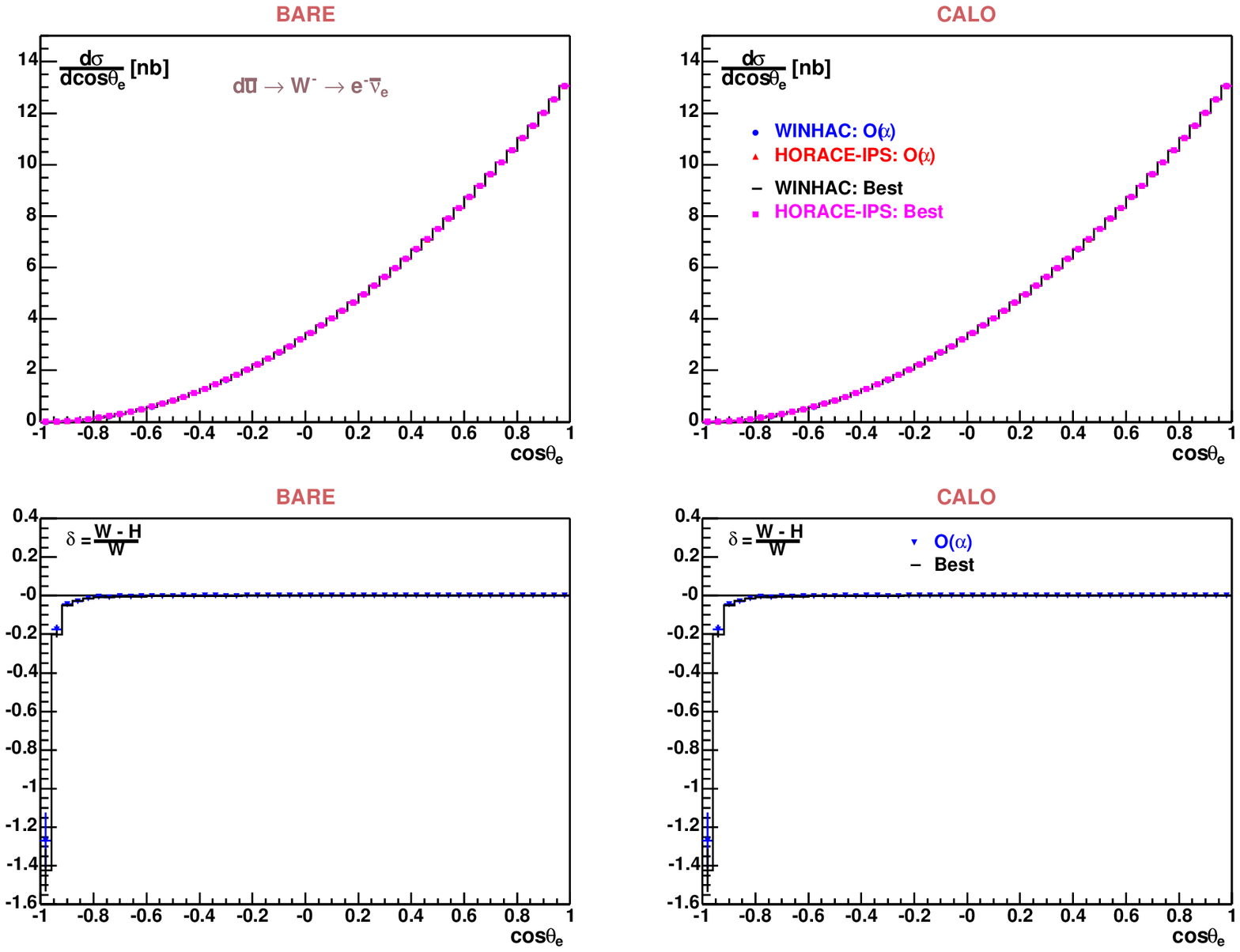,width=168mm,height=88mm}
}}
\end{picture}
\caption{\small\sf
Distributions of the electron polar angle at parton level
for \oal\ and ``Best'' predictions 
from \horace\ and \winhac, as well as their differences. The results are shown
for BARE (left) and CALO (right) event selections.
}
\label{fig:Par_el_costhl}
\end{figure}

\begin{figure}[!ht]
\setlength{\unitlength}{1mm}
\begin{picture}(160,88)
\put( -3,0){\makebox(0,0)[lb]{
\epsfig{file=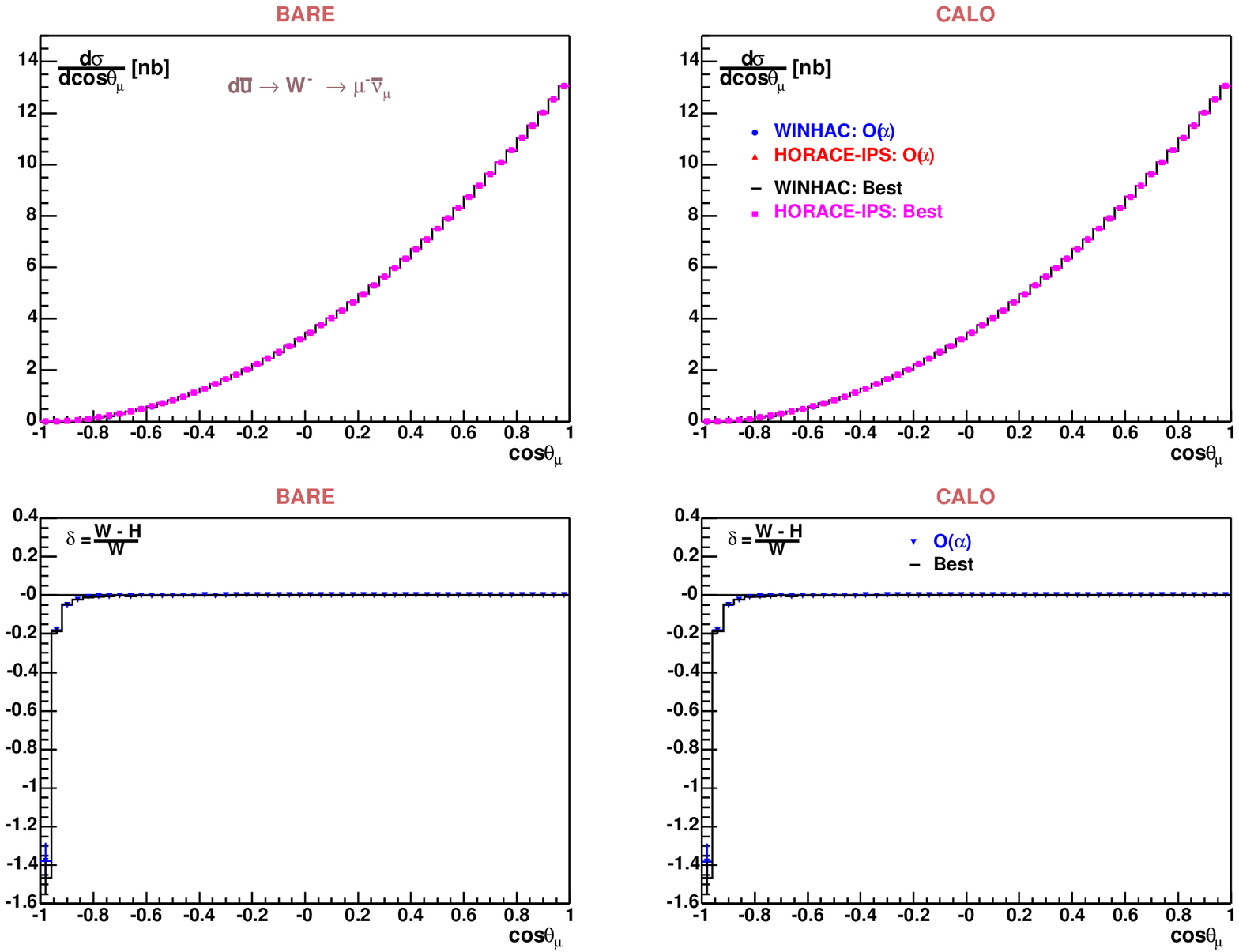,width=168mm,height=88mm}
}}
\end{picture}
\caption{\small\sf
Distributions of the muon polar angle at parton level
for \oal\ and ``Best'' predictions from 
\horace\ and \winhac, as well as their differences. The results are shown
for BARE (left) and CALO (right) event selections.
}
\label{fig:Par_mu_costhl}
\end{figure}

\begin{figure}[!ht]
\setlength{\unitlength}{1mm}
\begin{picture}(160,88)
\put( -3,0){\makebox(0,0)[lb]{
\epsfig{file=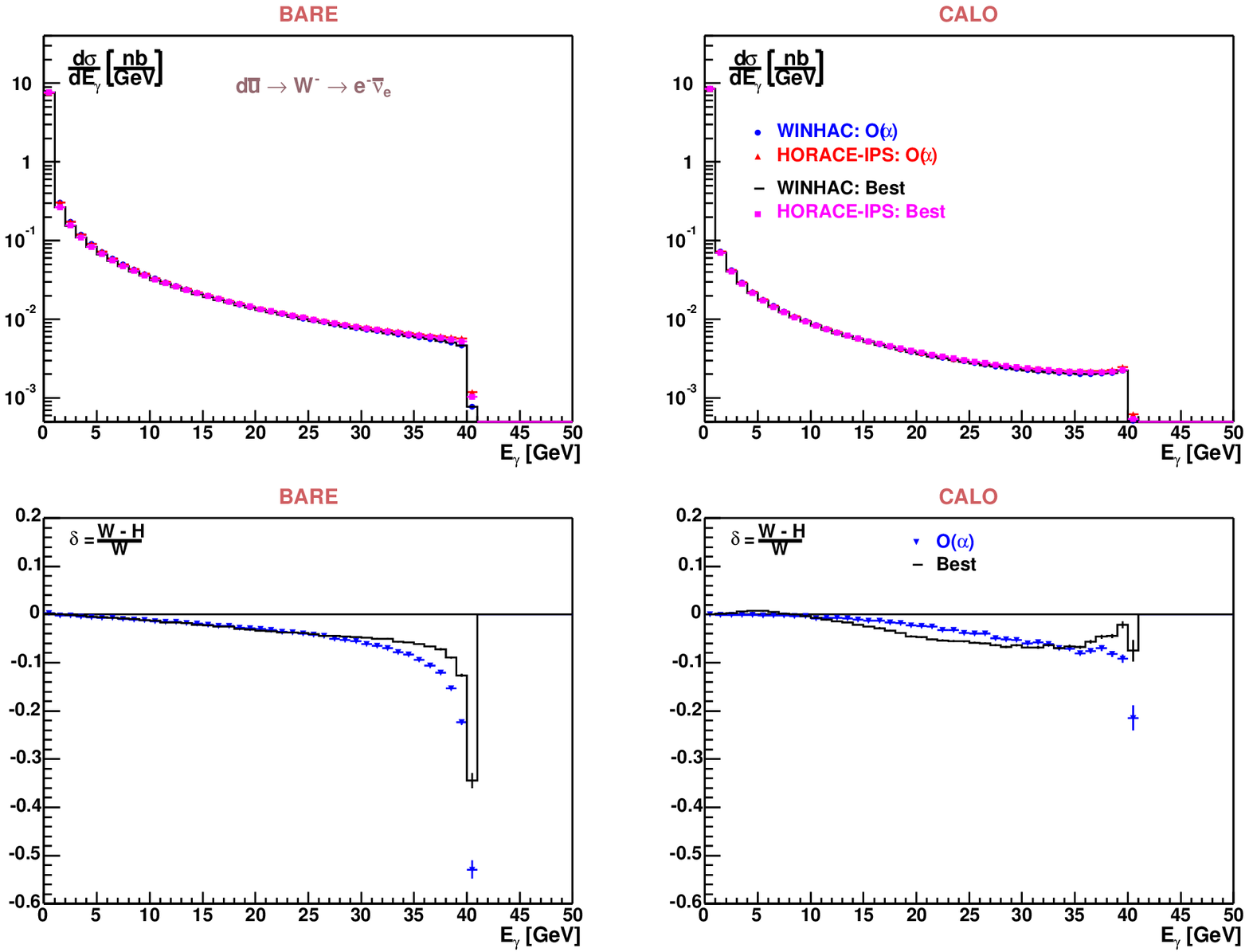,width=168mm,height=88mm}
}}
\end{picture}
\caption{\small\sf
Distributions of the hardest-photon energy at parton level in the electron
channel for \oal\ and ``Best'' predictions 
from \horace\ and \winhac, as well as their differences. 
The results are shown for BARE (left) and CALO (right) event selections.
}
\label{fig:Par_el_Eg}
\end{figure}

\begin{figure}[!ht]
\setlength{\unitlength}{1mm}
\begin{picture}(160,88)
\put( -3,0){\makebox(0,0)[lb]{
\epsfig{file=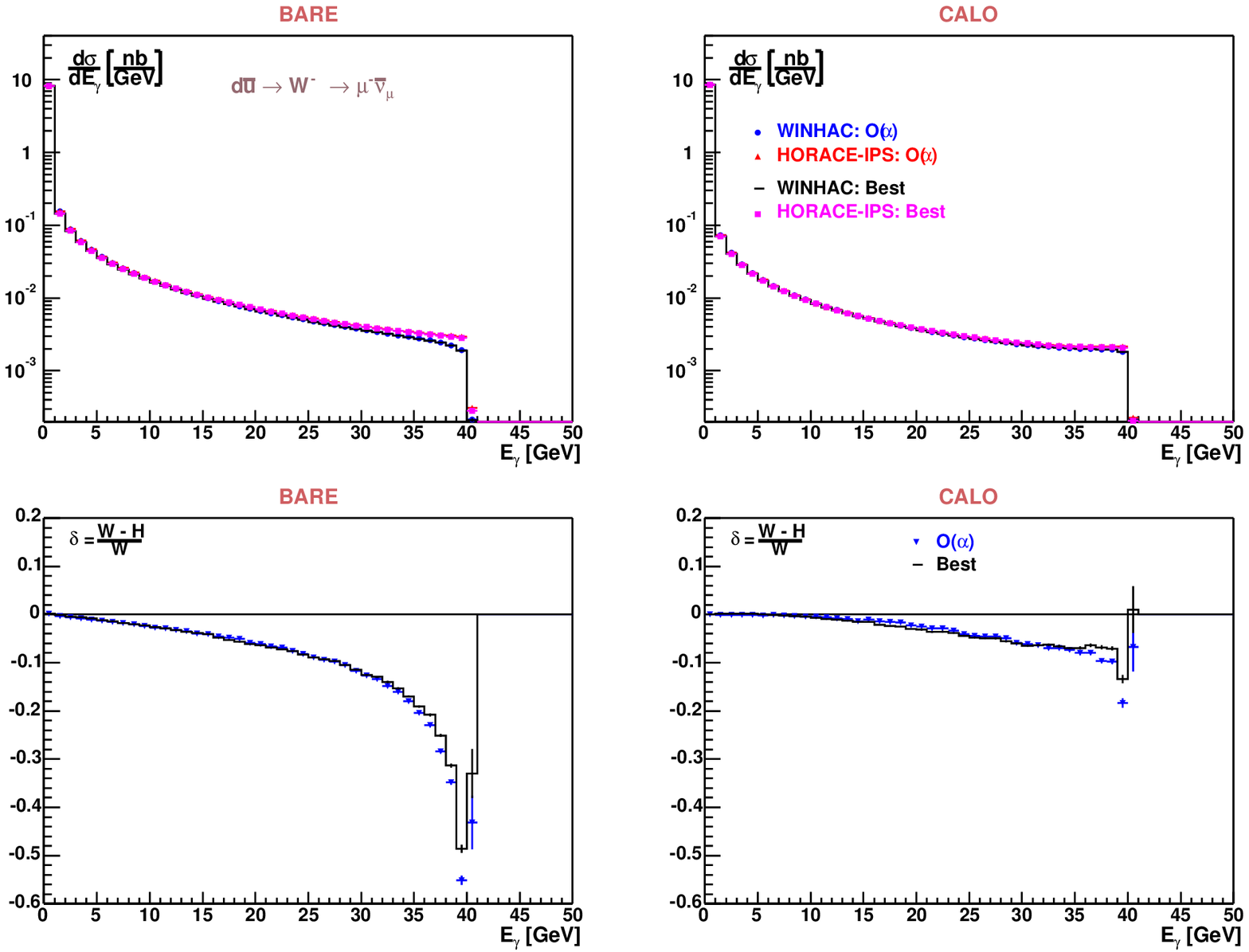,width=168mm,height=88mm}
}}
\end{picture}
\caption{\small\sf
Distributions of the hardest-photon energy at parton level in the muon channel 
for \oal\ and ``Best'' predictions from \horace\ and \winhac, as well as their
differences. 
The results are shown for BARE (left) and CALO (right) event selections.
}
\label{fig:Par_mu_Eg}
\end{figure}

\begin{figure}[!ht]
\setlength{\unitlength}{1mm}
\begin{picture}(160,88)
\put( -3,0){\makebox(0,0)[lb]{
\epsfig{file=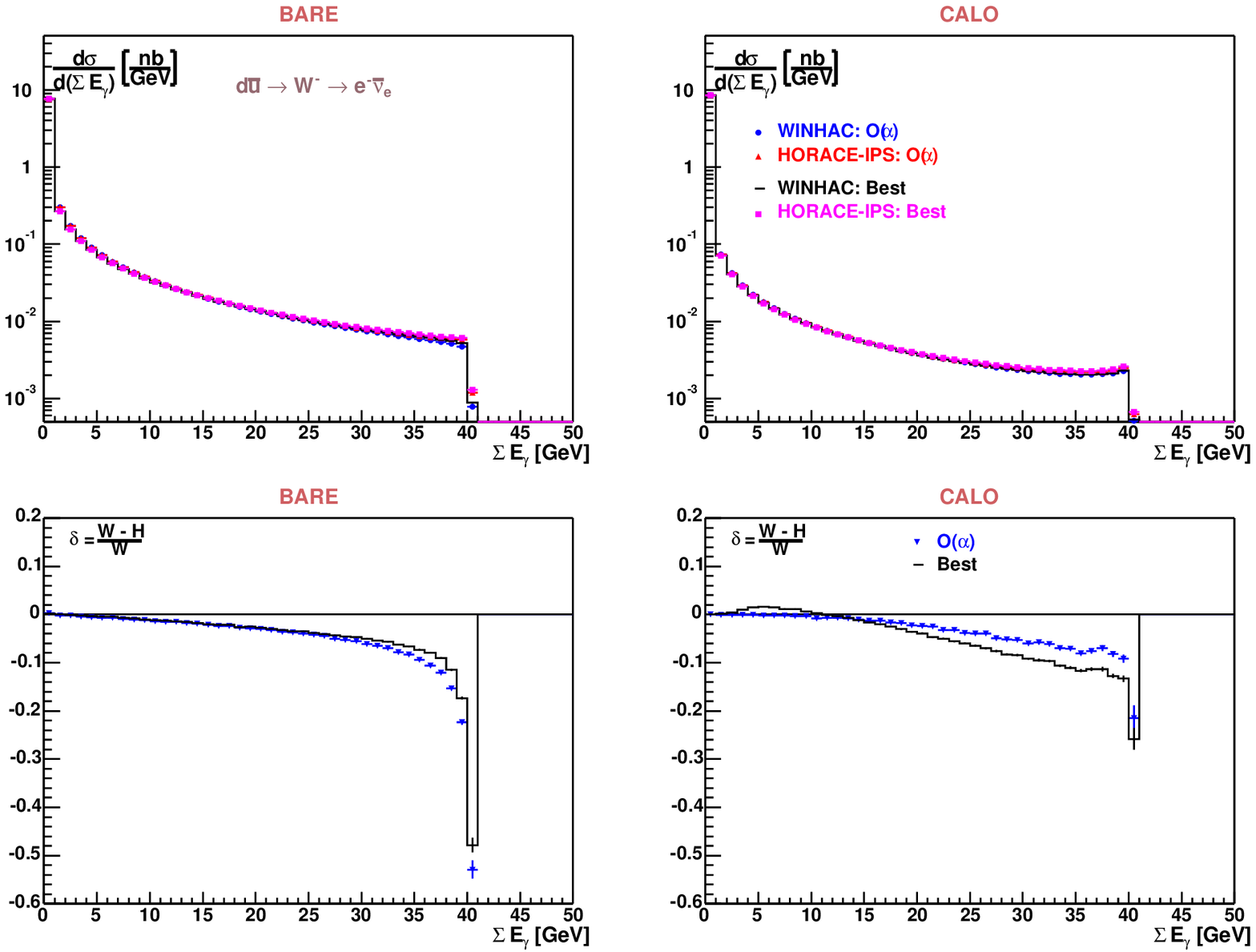,width=168mm,height=88mm}
}}
\end{picture}
\caption{\small\sf
Distributions of the total energy radiated by photons in the electron channel 
for \oal\ and ``Best'' predictions 
from \horace\ and \winhac, as well as their differences. 
The results are shown for BARE (left) and CALO (right) event selections.
}
\label{fig:Par_el_Egtot}
\end{figure}

\begin{figure}[!ht]
\setlength{\unitlength}{1mm}
\begin{picture}(160,88)
\put( -3,0){\makebox(0,0)[lb]{
\epsfig{file=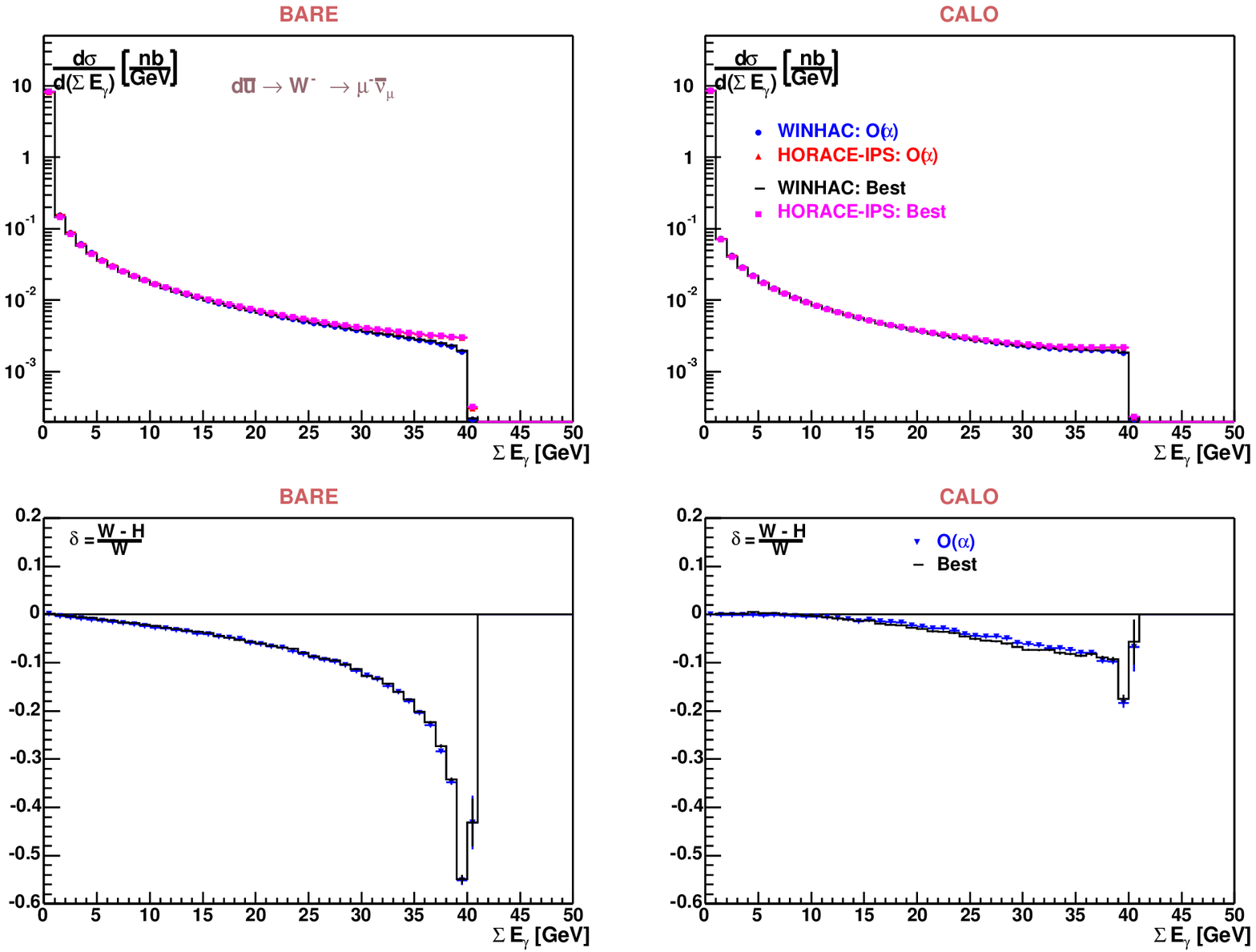,width=168mm,height=88mm}
}}
\end{picture}
\caption{\small\sf
Distributions of the total energy radiated by photons in the muon channel 
for \oal\ and ``Best'' predictions from \horace\ and \winhac, as well as their 
differences. 
The results are shown for BARE (left) and CALO (right) event selections.
}
\label{fig:Par_mu_Egtot}
\end{figure}

\begin{figure}[!ht]
\setlength{\unitlength}{1mm}
\begin{picture}(160,88)
\put( -3,0){\makebox(0,0)[lb]{
\epsfig{file=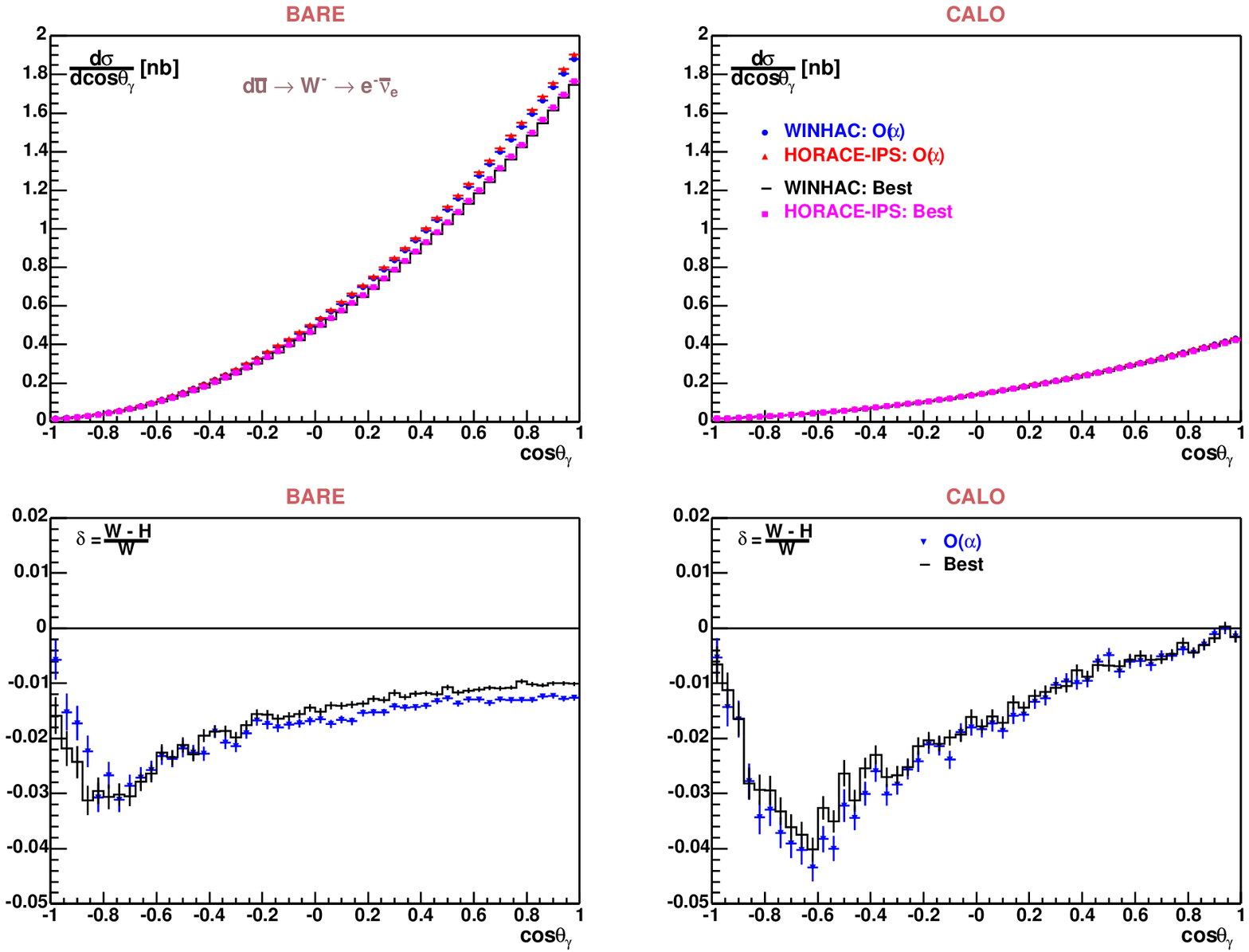,width=168mm,height=88mm}
}}
\end{picture}
\caption{\small\sf
Distributions of the hardest-photon polar angle in the electron channel 
for \oal\ and ``Best'' predictions from 
\horace\ and \winhac, as well as their differences. The results are shown
for BARE (left) and CALO (right) event selections.
}
\label{fig:Par_el_costhg}
\end{figure}

\begin{figure}[!ht]
\setlength{\unitlength}{1mm}
\begin{picture}(160,88)
\put( -3,0){\makebox(0,0)[lb]{
\epsfig{file=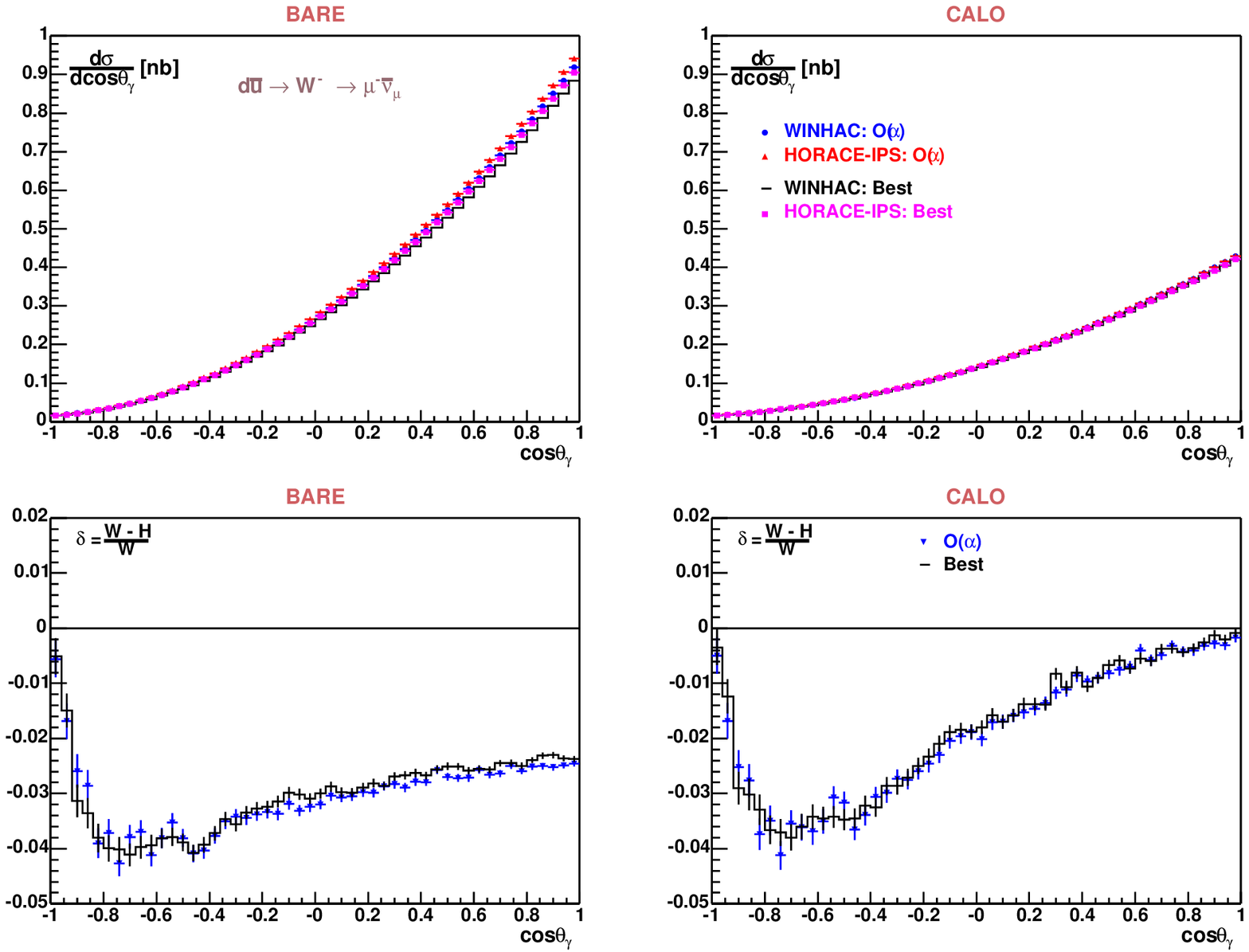,width=168mm,height=88mm}
}}
\end{picture}
\caption{\small\sf
Distributions of the hardest-photon polar angle in the muon channel
for \oal\ and ``Best'' predictions from 
\horace\ and \winhac, as well as their differences. 
The results are shown for BARE (left) and CALO (right) event selections.
}
\label{fig:Par_mu_costhg}
\end{figure}

%% file: haddif.tex
%
\begin{figure}[!ht]
\setlength{\unitlength}{1mm}
\begin{picture}(160,83)
\put( -3,-2){\makebox(0,0)[lb]{
\epsfig{file=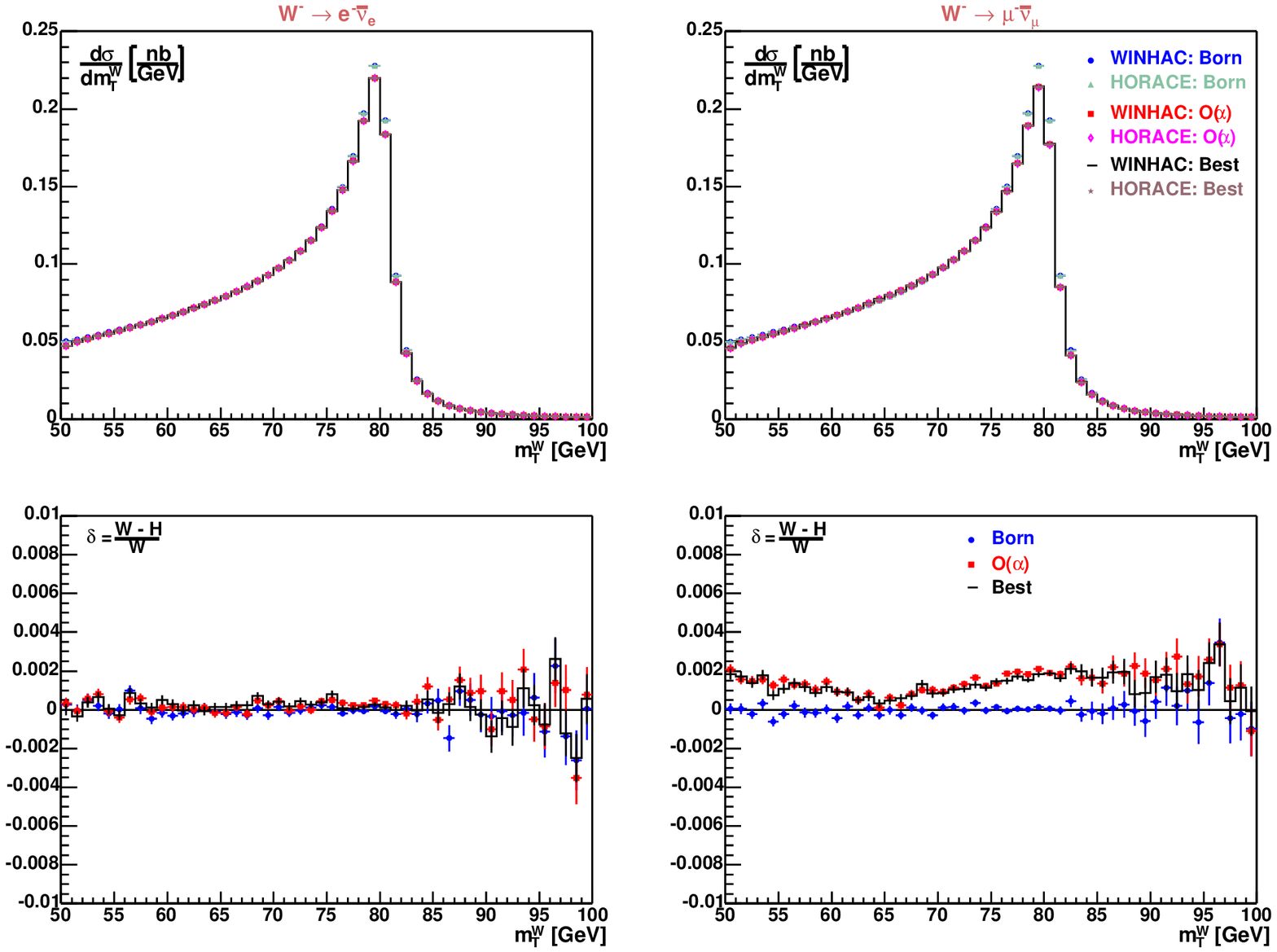,width=168mm,height=88mm}
}}
\end{picture}
\caption{\small\sf
Distributions of the $W^-$ transverse mass at hadron level for Born,
\oal\ and ``Best'' 
predictions from \horace\ and \winhac, as well as their differences. 
The results are shown for the electron (left) and muon (right) channels.
}
\label{fig:Had_Wm_MTW}
\end{figure}
%
\begin{figure}[!ht]
\setlength{\unitlength}{1mm}
\begin{picture}(160,83)
\put( -3,-2){\makebox(0,0)[lb]{
\epsfig{file=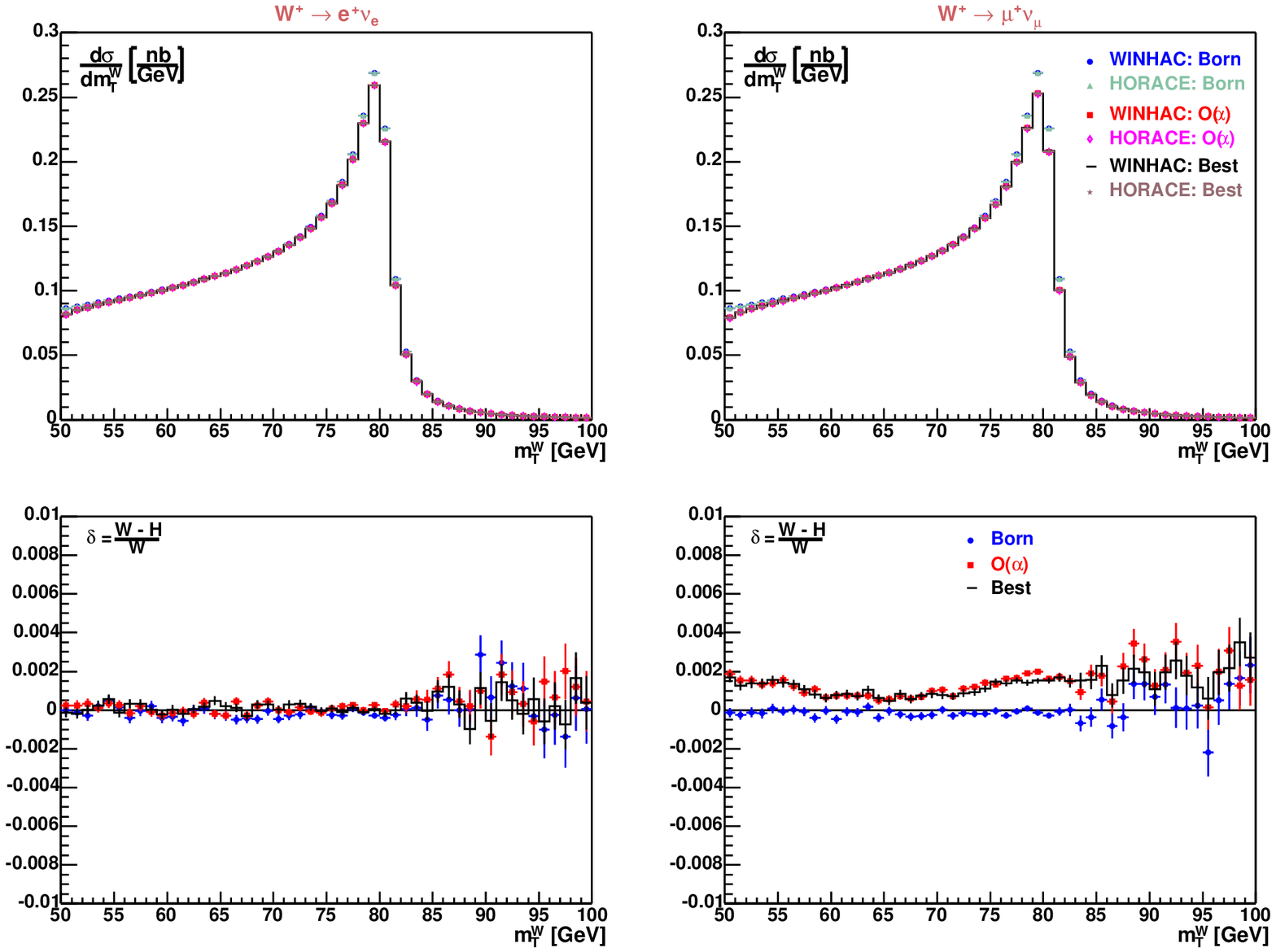,width=168mm,height=88mm}
}}
\end{picture}
\caption{\small\sf
Distributions of the $W^+$ transverse mass at hadron level 
for Born, \oal\ and ``Best'' 
predictions from \horace\ and \winhac, as well as their differences. 
The results are shown for the electron (left) and muon (right) channels.
}
\label{fig:Had_Wp_MTW}
\end{figure}

\begin{figure}[!ht]
\setlength{\unitlength}{1mm}
\begin{picture}(160,87)
\put( -3,-1){\makebox(0,0)[lb]{
\epsfig{file=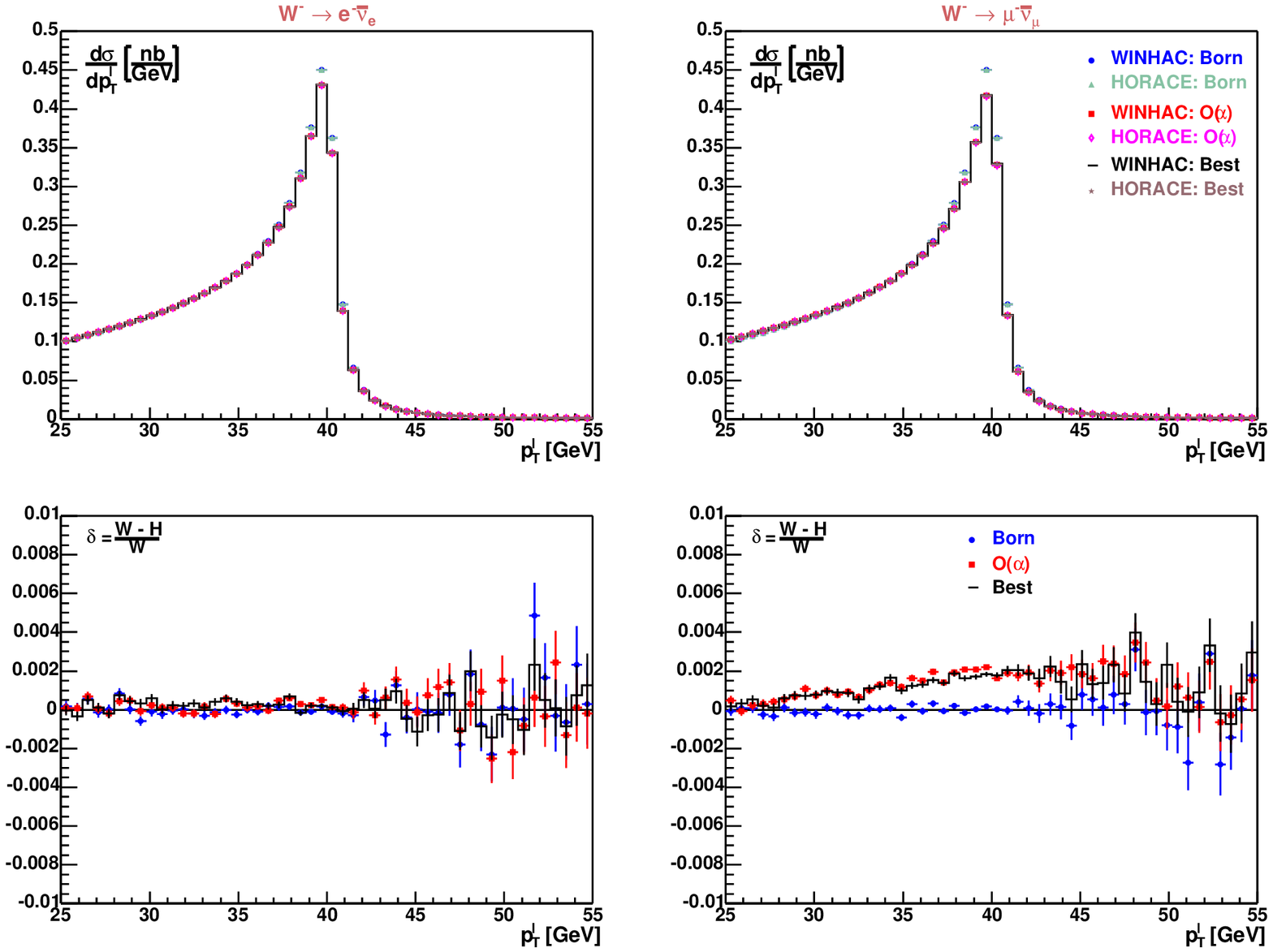,width=168mm,height=88mm}
}}
\end{picture}
\caption{\small\sf
Distributions of the charged lepton transverse momentum at hadron level
for Born, 
\oal\ and ``Best'' predictions of $W^-$ production 
from \horace\ and \winhac, as well as their differences. 
The results are shown for the electron (left) and muon (right) channels.
}
\label{fig:Had_Wm_pTl}
\end{figure}

\begin{figure}[!ht]
\setlength{\unitlength}{1mm}
\begin{picture}(160,87)
\put( -3,-1){\makebox(0,0)[lb]{
\epsfig{file=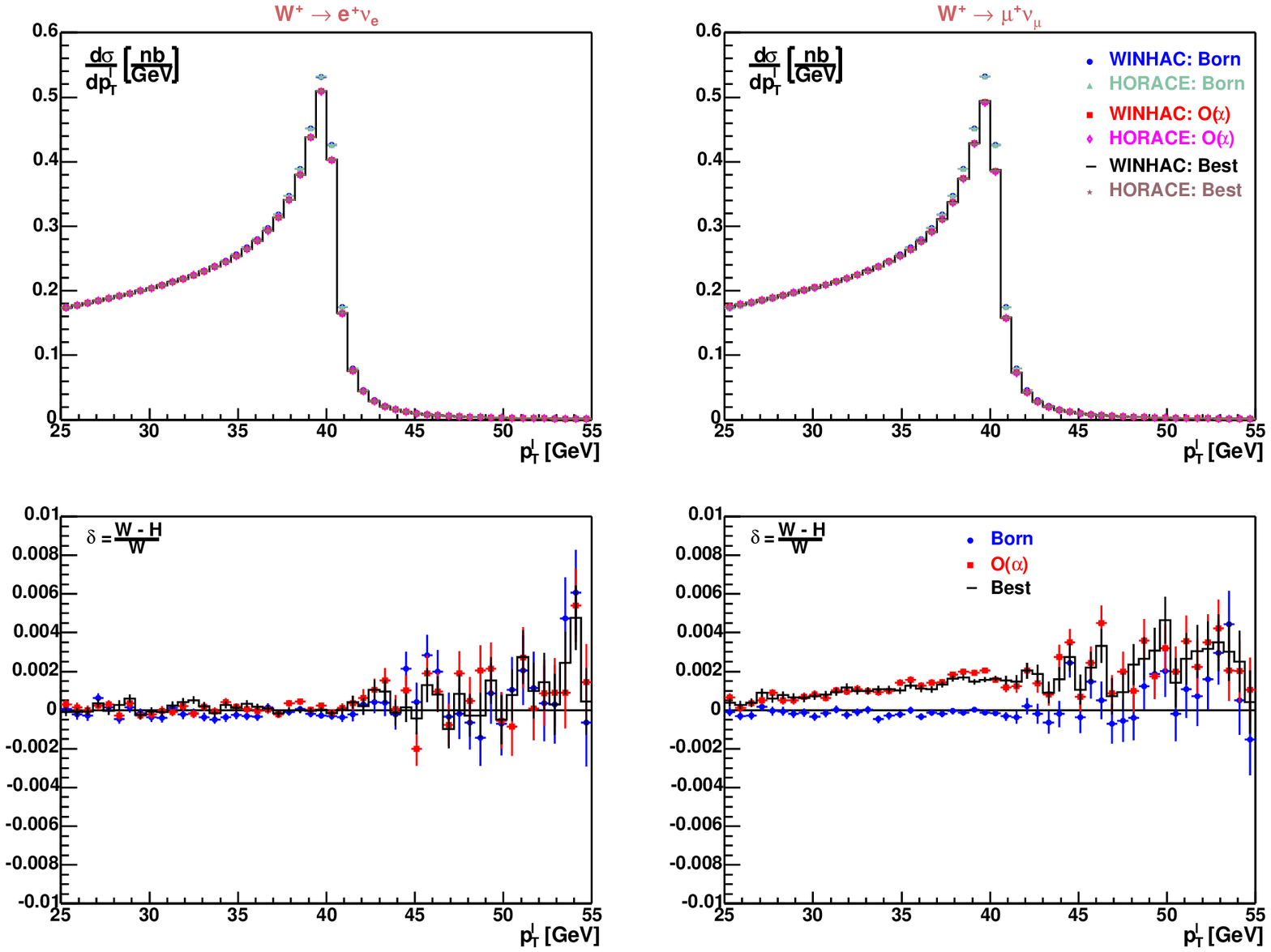,width=168mm,height=88mm}
}}
\end{picture}
\caption{\small\sf
Distributions of the charged lepton transverse momentum at hadron level
for Born, 
\oal\ and ``Best'' predictions of $W^+$ production
from \horace\ and \winhac, as well as their differences. 
The results are shown for the electron (left) and muon (right) channels.
}
\label{fig:Had_Wp_pTl}
\end{figure}

\begin{figure}[!ht]
\setlength{\unitlength}{1mm}
\begin{picture}(160,88)
\put( -3,0){\makebox(0,0)[lb]{
\epsfig{file=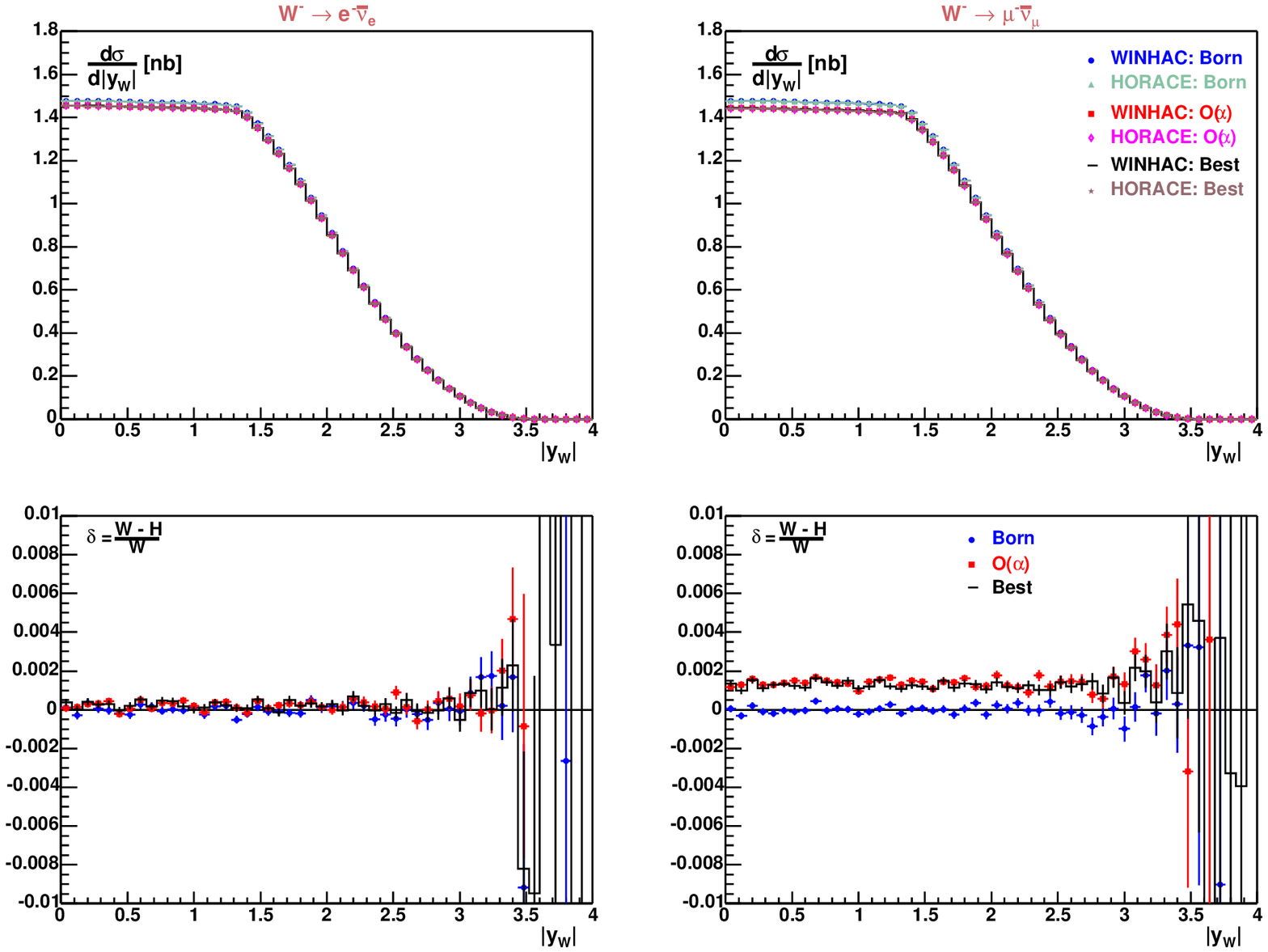,width=168mm,height=88mm}
}}
\end{picture}
\caption{\small\sf
Distributions of the $W^-$ rapidity at hadron level
for Born, \oal\ and ``Best'' 
predictions from \horace\ and \winhac, as well as their differences. 
The results are shown for the electron (left) and muon (right) channels.
}
\label{fig:Had_Wm_yW}
\end{figure}

\begin{figure}[!ht]
\setlength{\unitlength}{1mm}
\begin{picture}(160,88)
\put( -3,0){\makebox(0,0)[lb]{
\epsfig{file=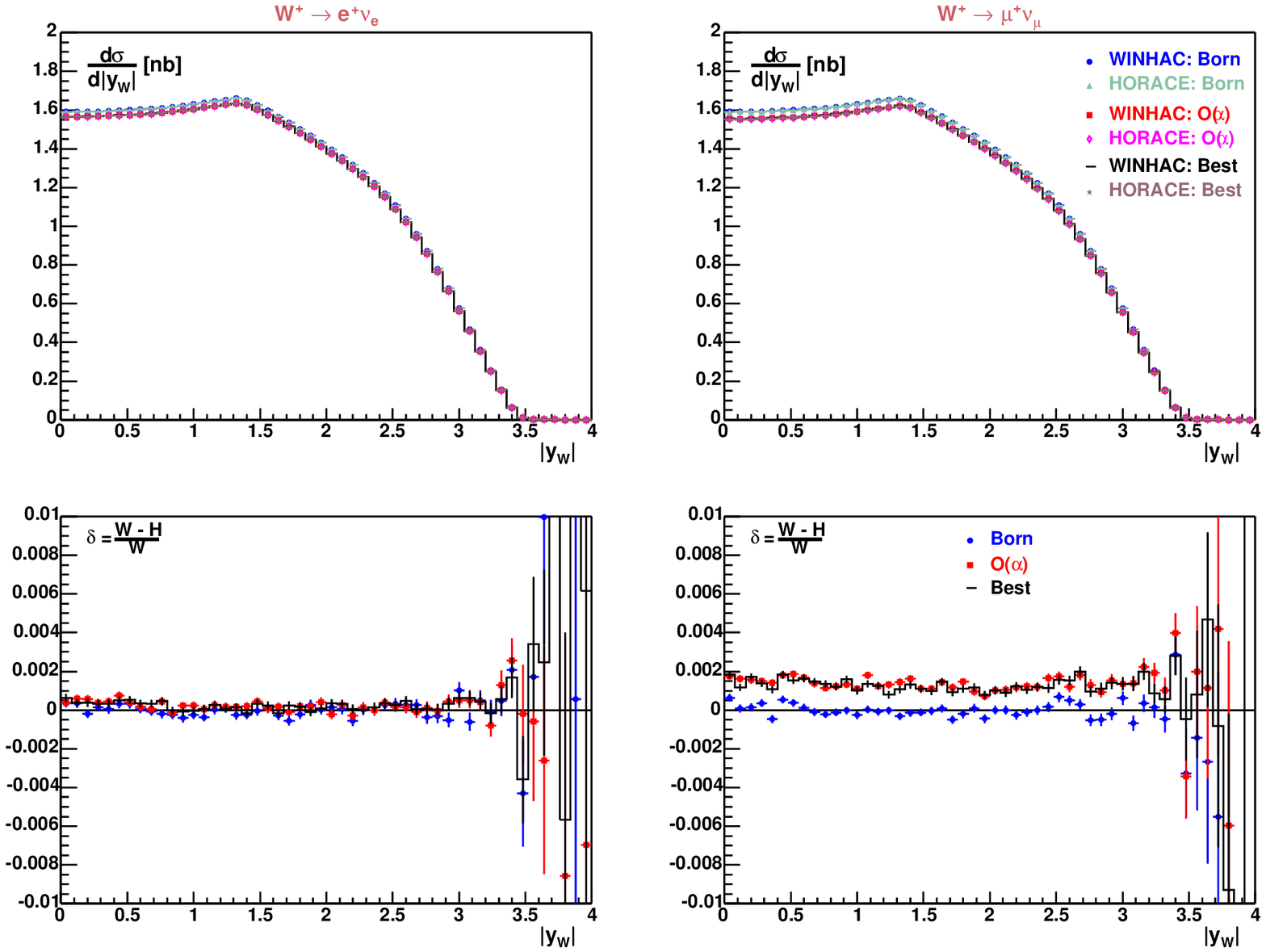,width=168mm,height=88mm}
}}
\end{picture}
\caption{\small\sf
Distributions of the $W^+$ rapidity at hadron level for Born, \oal\ and ``Best'' 
predictions from \horace\ and \winhac, as well as their differences. 
The results are shown for the electron (left) and muon (right) channels.
}
\label{fig:Had_Wp_yW}
\end{figure}

\begin{figure}[!ht]
\setlength{\unitlength}{1mm}
\begin{picture}(160,87)
\put( -3,-1){\makebox(0,0)[lb]{
\epsfig{file=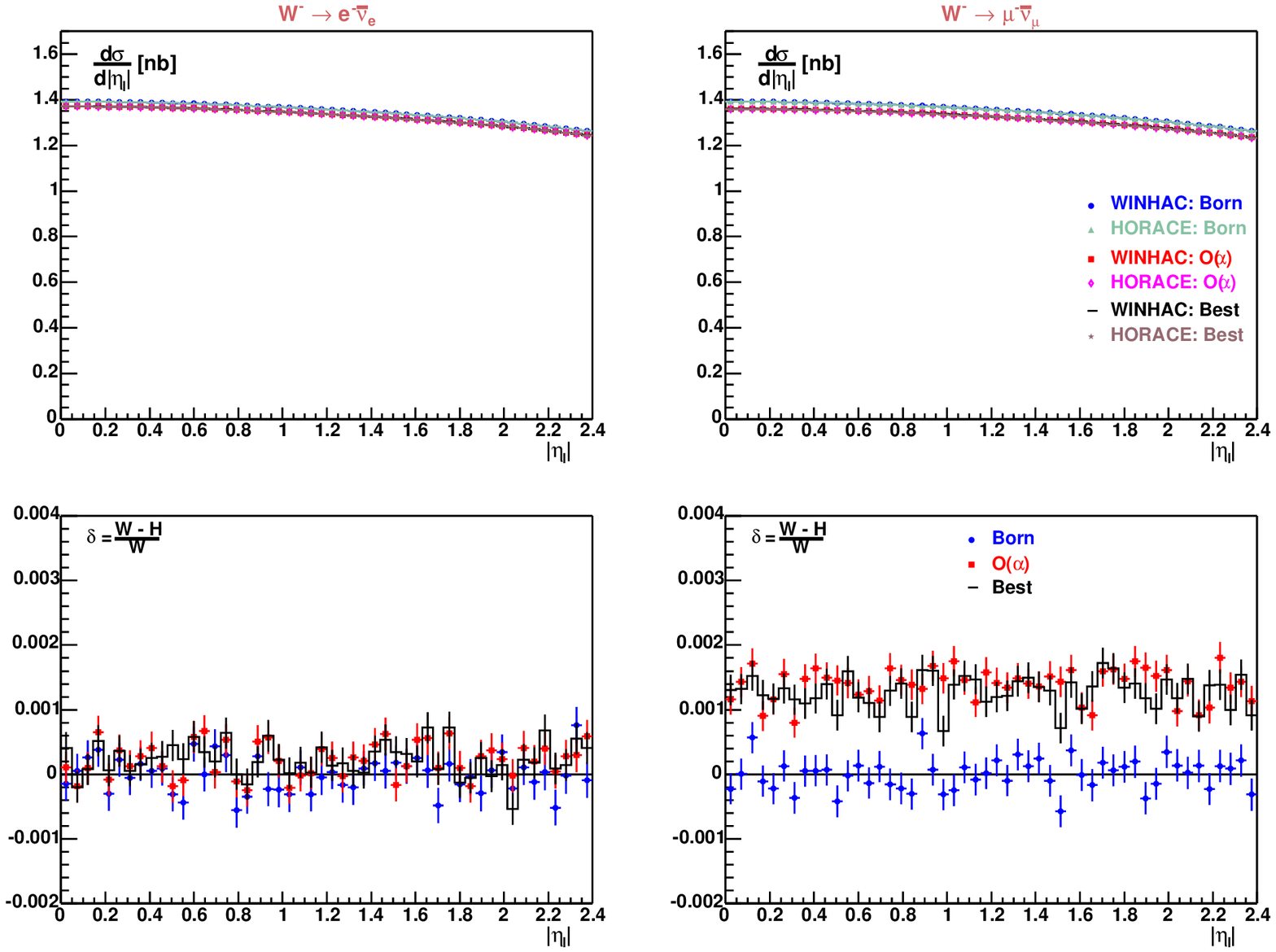,width=168mm,height=88mm}
}}
\end{picture}
\caption{\small\sf
Distributions of the charged lepton pseudorapidity at hadron level for Born, 
\oal\ and ``Best'' predictions of $W^-$ production 
from \horace\ and \winhac, as well as their differences. 
The results are shown for the electron (left) and muon (right) channels.
}
\label{fig:Had_Wm_etal}
\end{figure}

\begin{figure}[!ht]
\setlength{\unitlength}{1mm}
\begin{picture}(160,87)
\put( -3,-1){\makebox(0,0)[lb]{
\epsfig{file=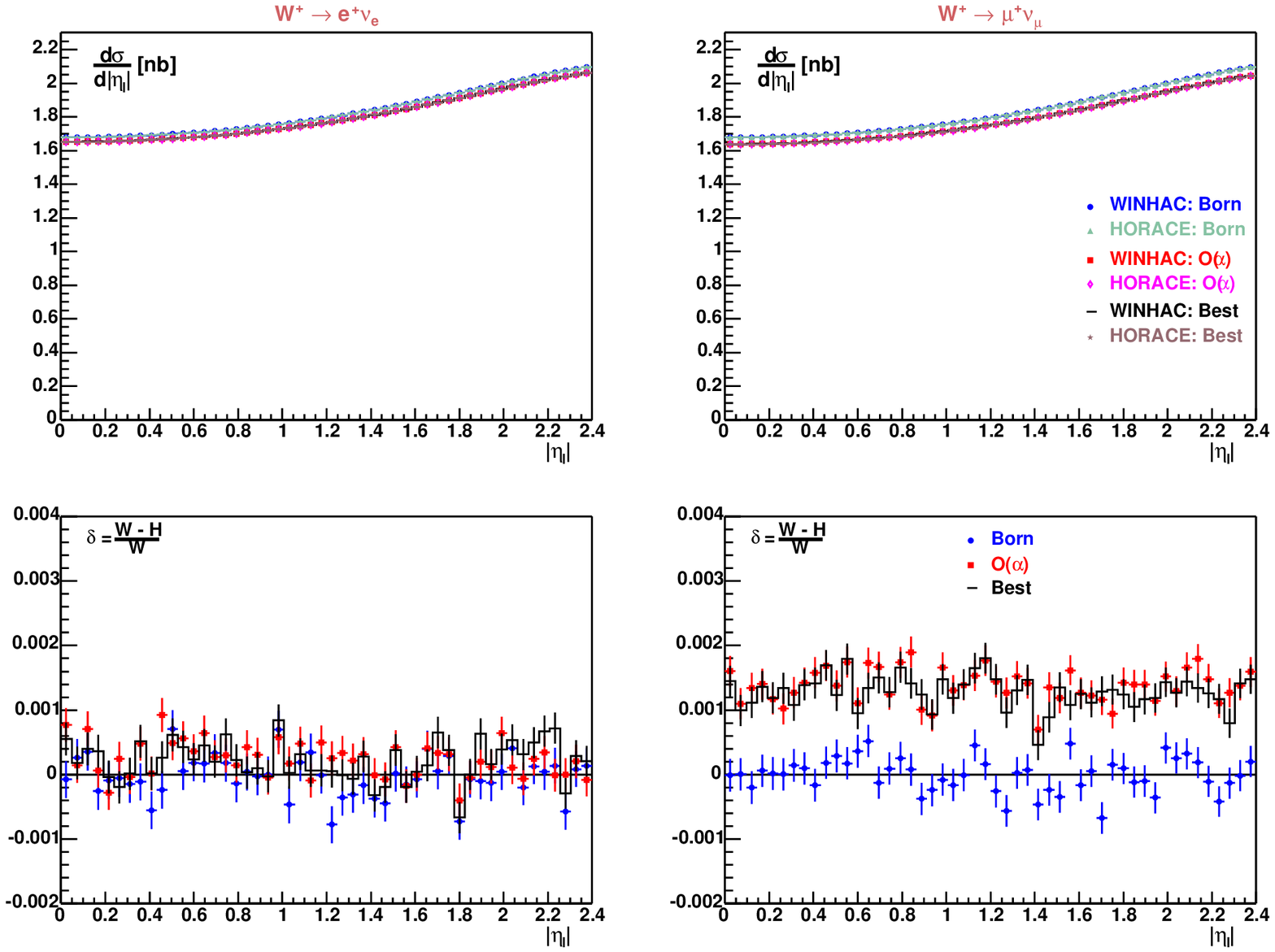,width=168mm,height=88mm}
}}
\end{picture}
\caption{\small\sf
Distributions of the charged lepton pseudorapidity at hadron level for Born, 
\oal\ and ``Best'' predictions of $W^+$ production 
from \horace\ and \winhac, as well as their differences. 
The results are shown for the electron (left) and muon (right) channels.
}
\label{fig:Had_Wp_etal}
\end{figure}

\begin{figure}[!ht]
\setlength{\unitlength}{1mm}
\begin{picture}(160,87)
\put( -3,-1){\makebox(0,0)[lb]{
\epsfig{file=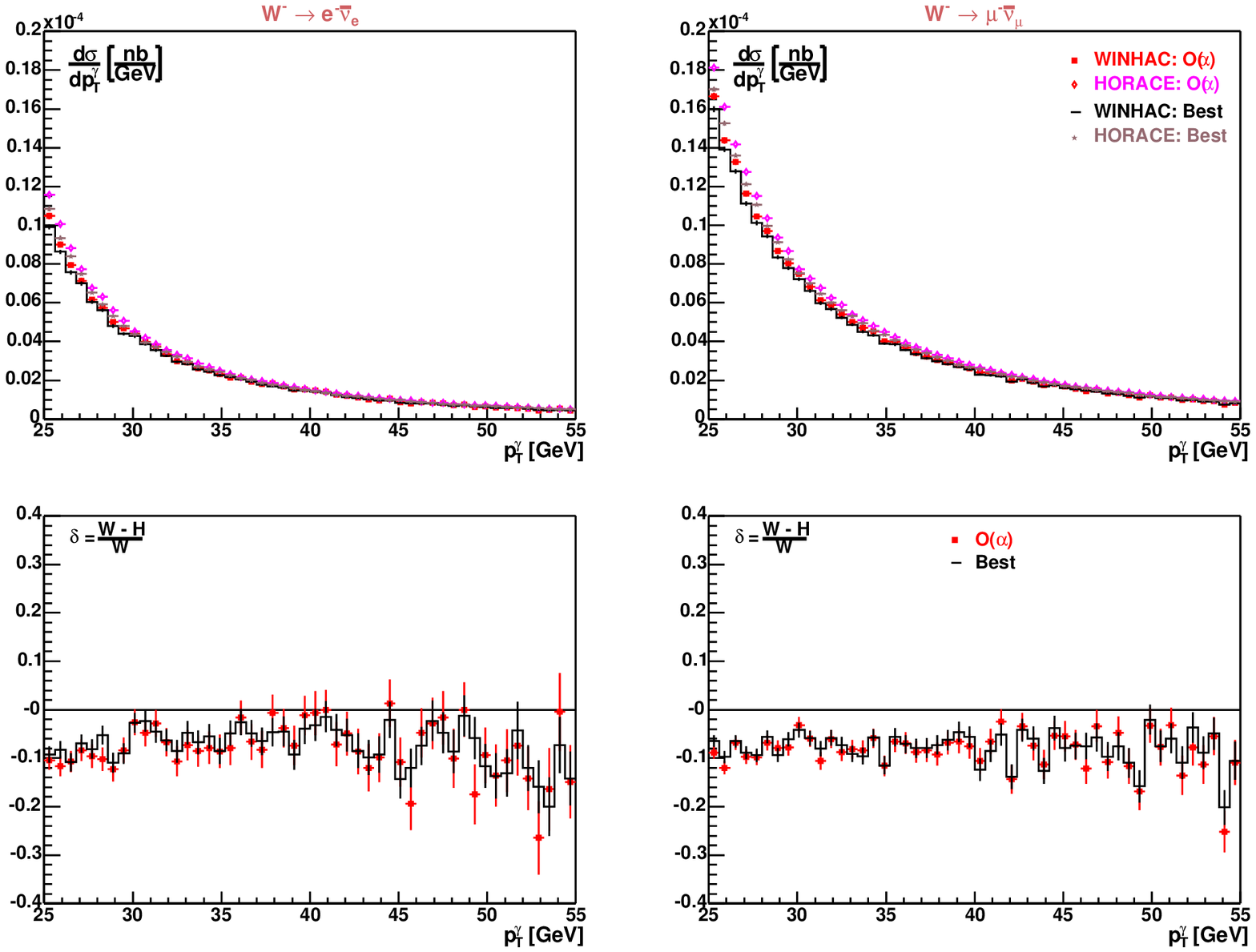,width=168mm,height=88mm}
}}
\end{picture}
\caption{\small\sf
Distributions of the hardest-photon transverse momentum at hadron level for  
\oal\ and ``Best'' predictions of $W^-$ production 
from \horace\ and \winhac, as well as their differences. 
The results are shown for the electron (left) and muon (right) channels.
}
\label{fig:Had_Wm_pTg}
\end{figure}

\begin{figure}[!ht]
\setlength{\unitlength}{1mm}
\begin{picture}(160,87)
\put( -3,-1){\makebox(0,0)[lb]{
\epsfig{file=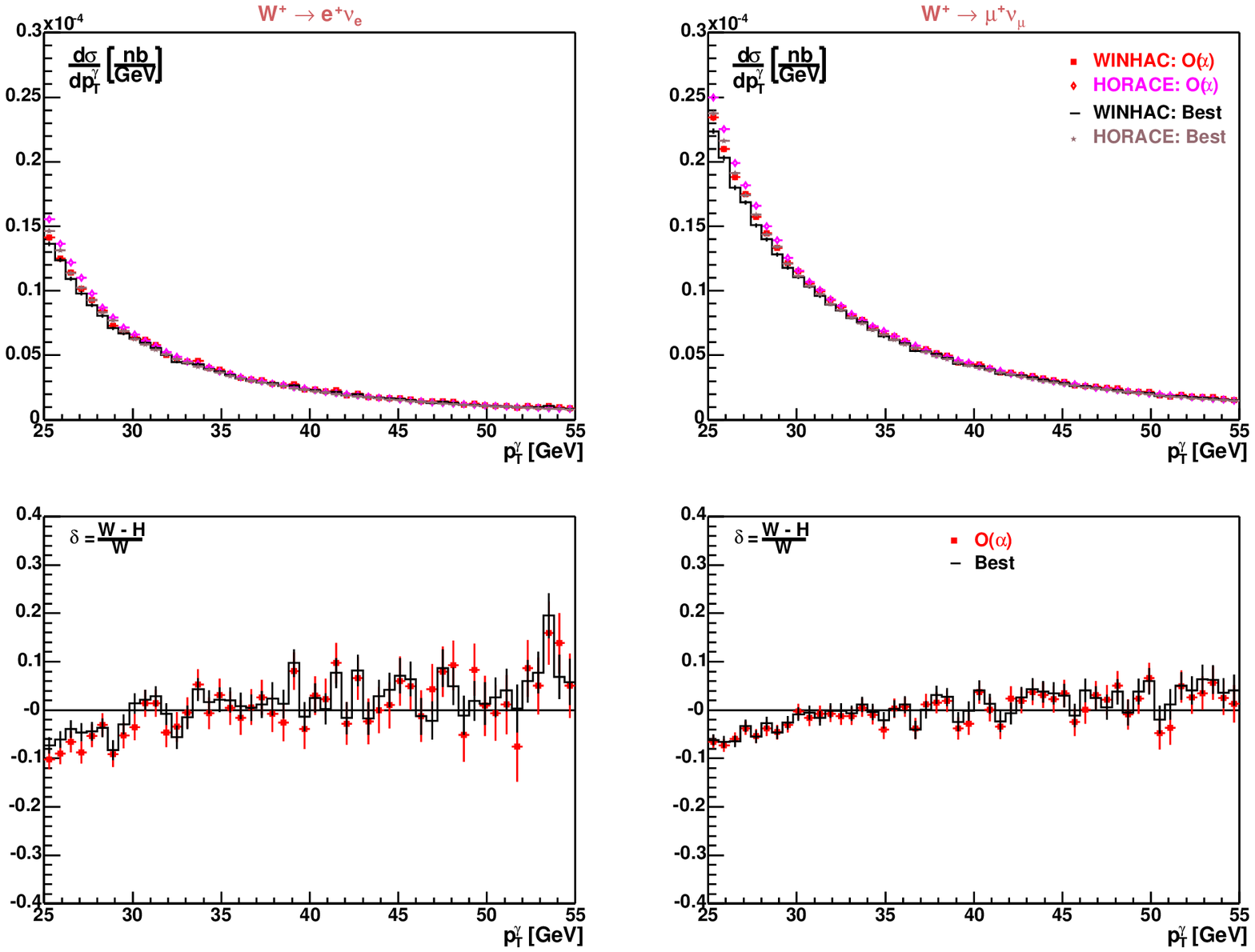,width=168mm,height=88mm}
}}
\end{picture}
\caption{\small\sf
Distributions of the hardest-photon transverse momentum at hadron level for  
\oal\ and ``Best'' predictions of $W^+$ production 
from \horace\ and \winhac, as well as their differences. 
The results are shown for the electron (left) and muon (right) channels.
}
\label{fig:Had_Wp_pTg}
\end{figure}

\begin{figure}[!ht]
\setlength{\unitlength}{1mm}
\begin{picture}(160,87)
\put( -3,-1){\makebox(0,0)[lb]{
\epsfig{file=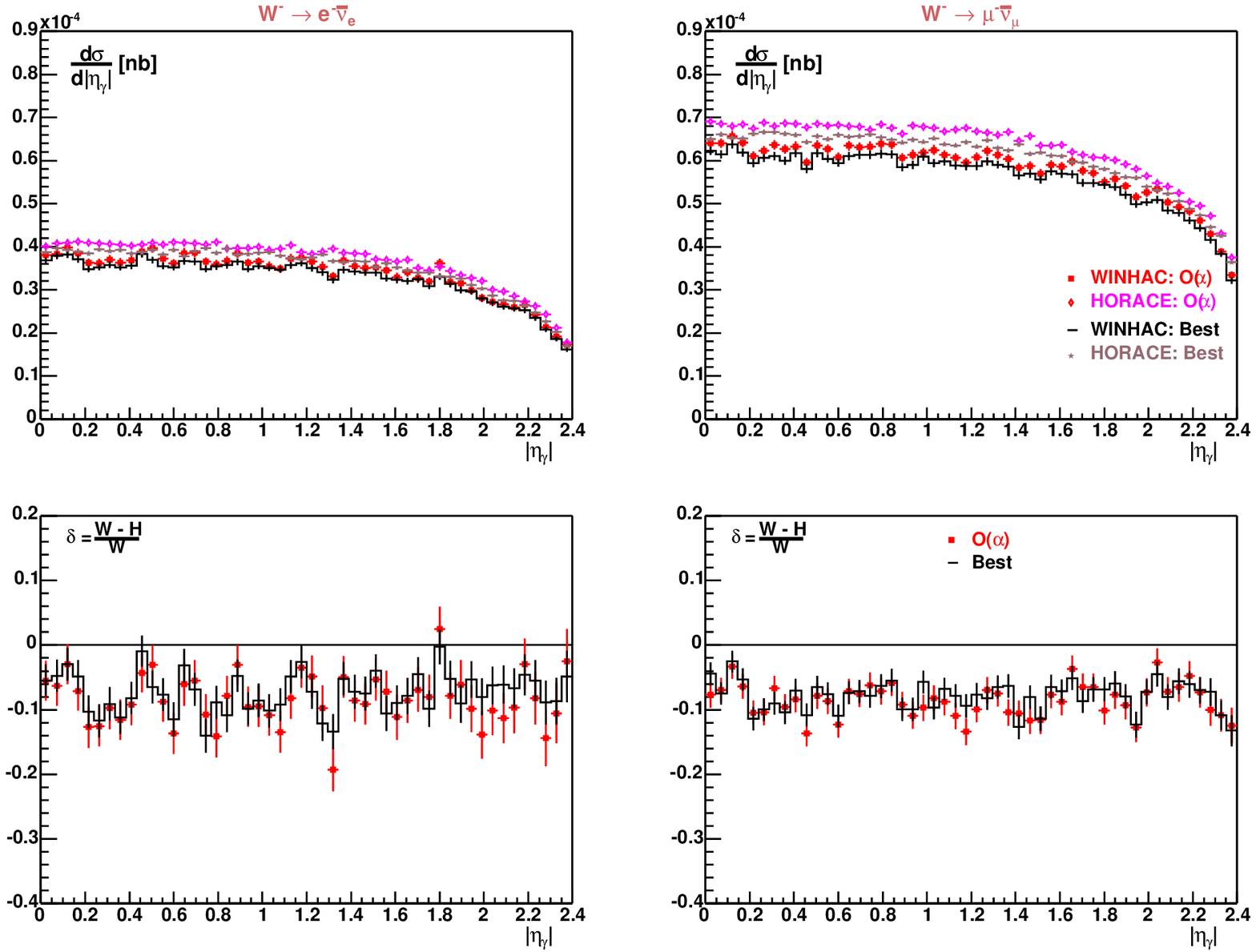,width=168mm,height=88mm}
}}
\end{picture}
\caption{\small\sf
Distributions of the hardest-photon pseudorapidity at hadron level for
\oal\ and ``Best'' predictions of $W^-$ production 
from \horace\ and \winhac, as well as their differences. 
The results are shown for the electron (left) and muon (right) channels.
}
\label{fig:Had_Wm_etag}
\end{figure}

\begin{figure}[!ht]
\setlength{\unitlength}{1mm}
\begin{picture}(160,87)
\put( -3,-1){\makebox(0,0)[lb]{
\epsfig{file=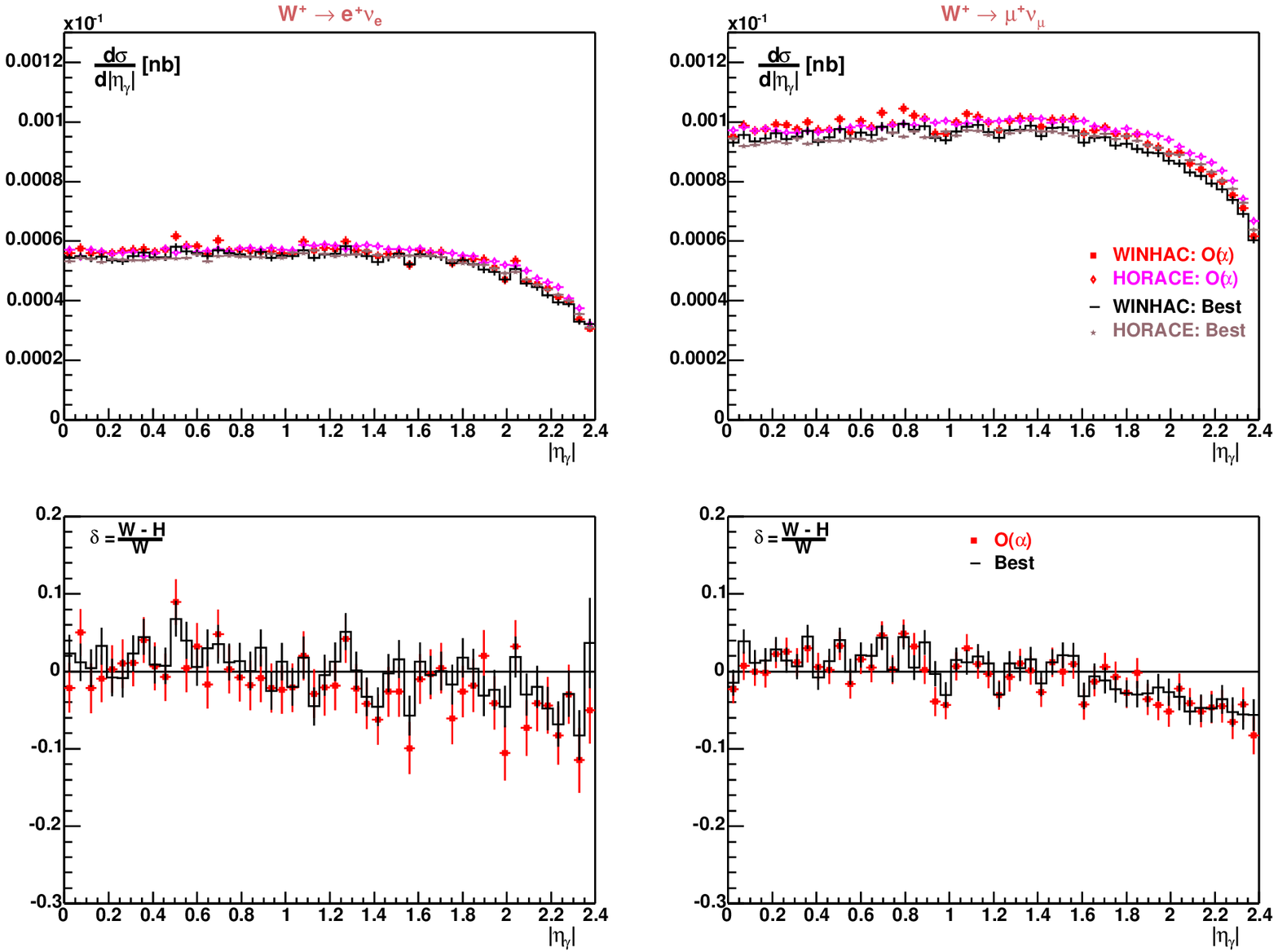,width=168mm,height=88mm}
}}
\end{picture}
\caption{\small\sf
Distributions of the hardest-photon pseudorapidity at hadron level for
\oal\ and ``Best'' predictions of $W^+$ production 
from \horace\ and \winhac, as well as their differences. 
The results are shown for the electron (left) and muon (right) channels.
}
\label{fig:Had_Wp_etag}
\end{figure}

%% file: hadcor.tex
%
\begin{figure}[!ht]
\setlength{\unitlength}{1mm}
\begin{picture}(160,75)
\put( -3,-2){\makebox(0,0)[lb]{
\epsfig{file=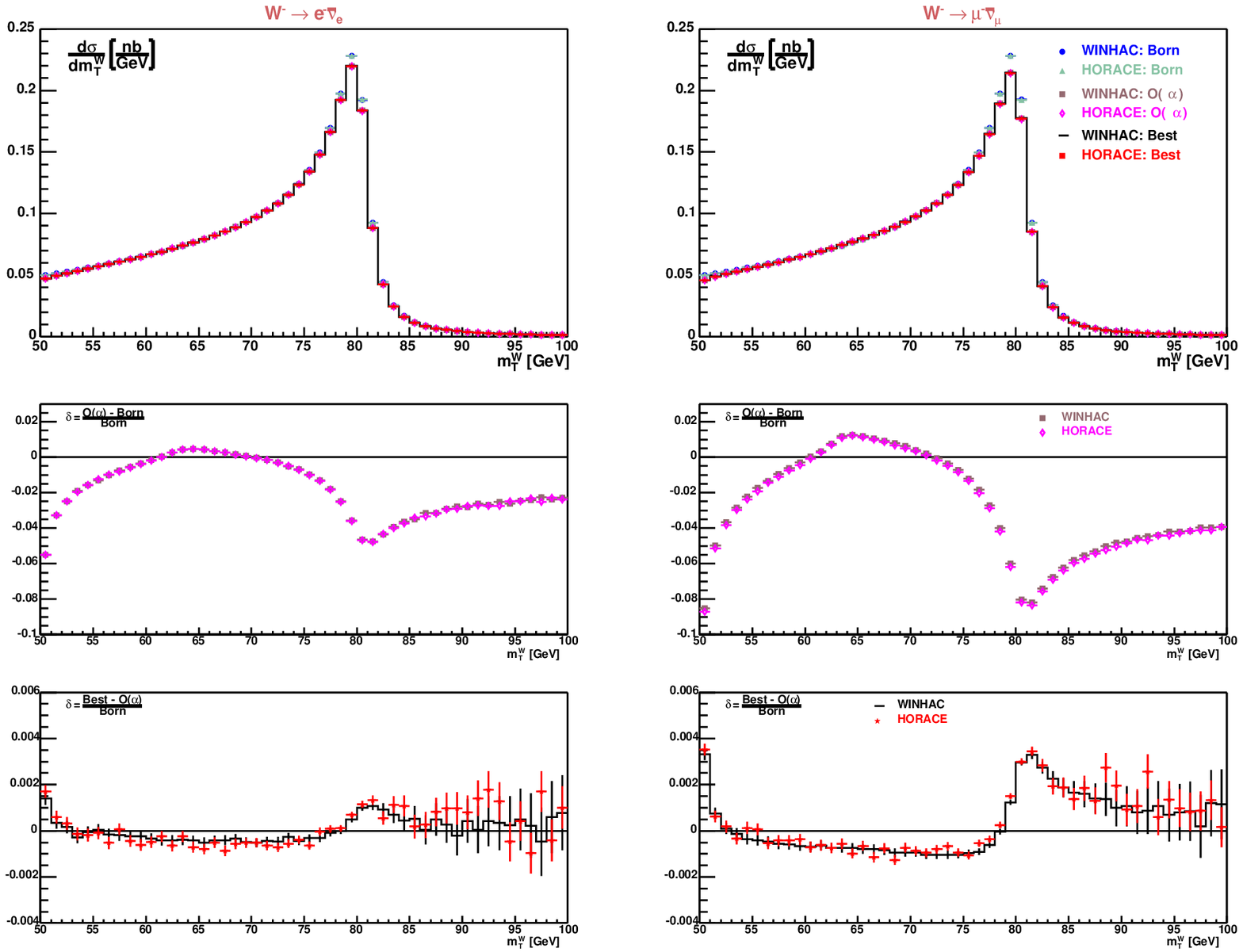,width=168mm,height=88mm}
}}
\end{picture}
\caption{\small\sf
Distributions of the $W^-$ transverse mass at hadron level
for Born, \oal\ and ``Best'' 
predictions from \horace\ and \winhac, and the size of QED corrections. 
The results are shown for the electron (left) and muon (right) channels.
}
\label{fig:Had_Wm_MTW_Cor}
\end{figure}
%

\begin{figure}[!ht]
\setlength{\unitlength}{1mm}
\begin{picture}(160,87)
\put( -3,-1){\makebox(0,0)[lb]{
\epsfig{file=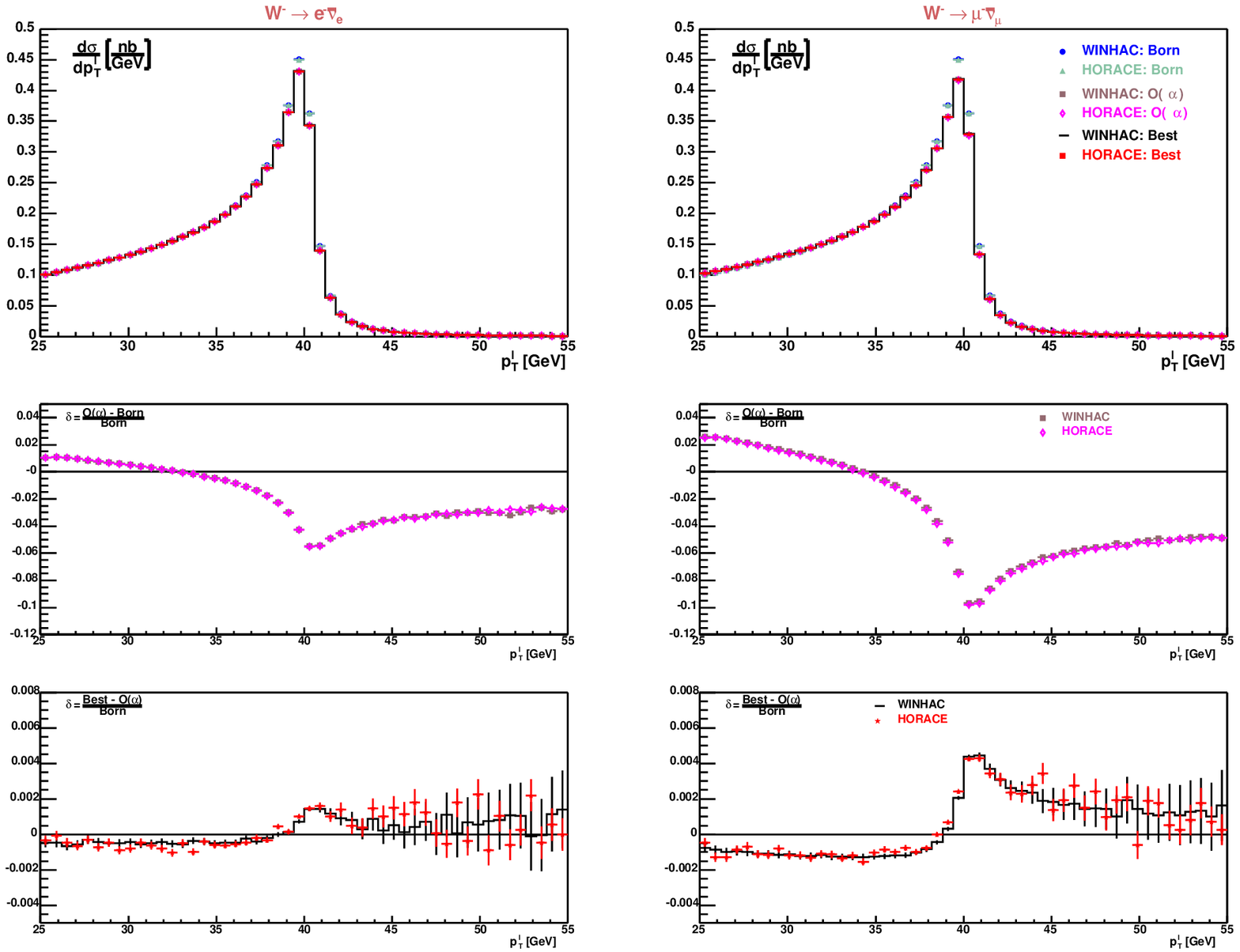,width=168mm,height=88mm}
}}
\end{picture}
\caption{\small\sf
Distributions of the charged lepton transverse momentum at hadron level
for Born, \oal\ and ``Best'' predictions of $W^-$ production 
from \horace\ and \winhac, and the size of QED corrections. 
The results are shown for the electron (left) and muon (right) channels.
}
\label{fig:Had_Wm_pTl_Cor}
\end{figure}


\begin{figure}[!ht]
\setlength{\unitlength}{1mm}
\begin{picture}(160,88)
\put( -3,0){\makebox(0,0)[lb]{
\epsfig{file=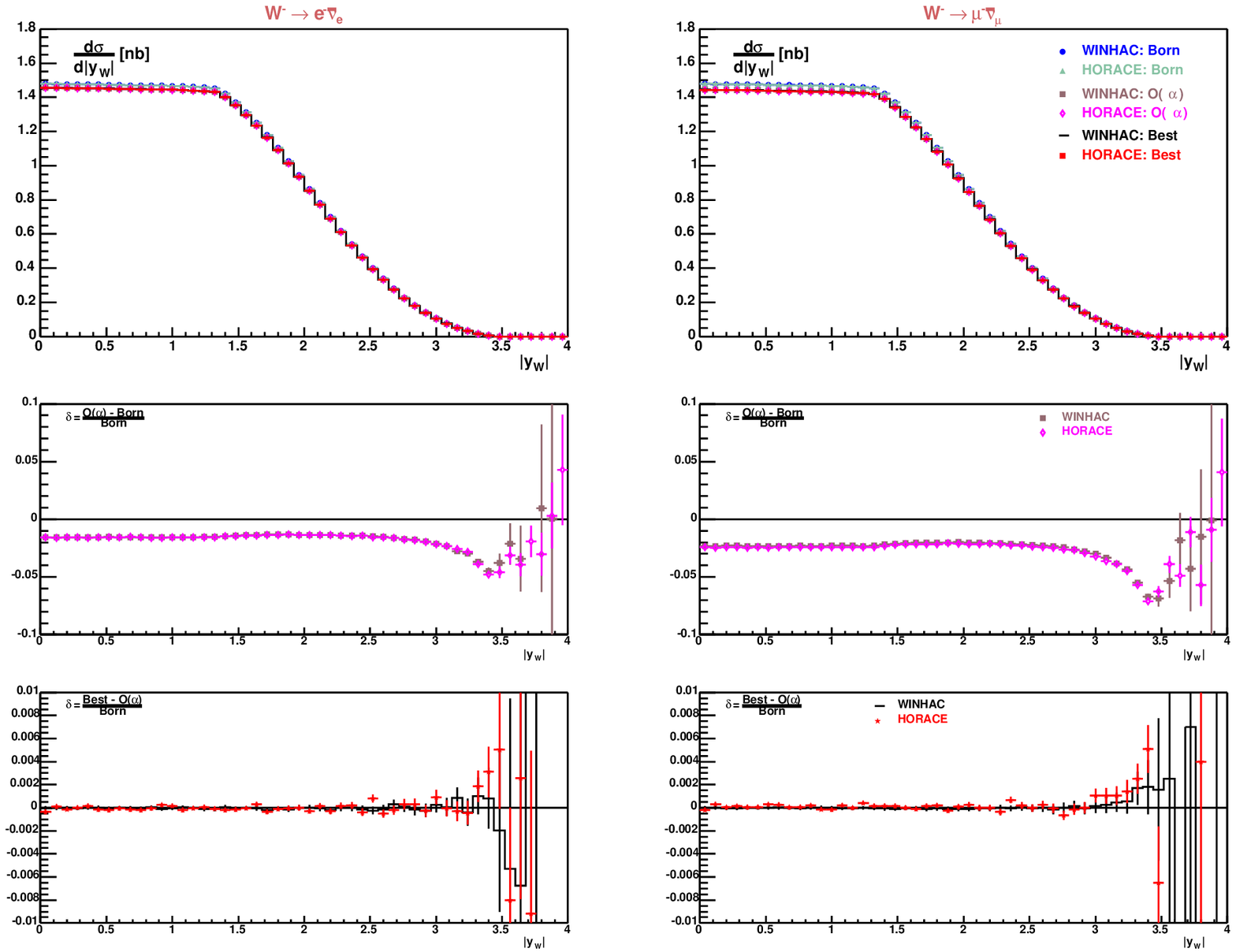,width=168mm,height=88mm}
}}
\end{picture}
\caption{\small\sf
Distributions of the $W^-$ rapidity at hadron level for Born, \oal\ and ``Best'' 
predictions from \horace\ and \winhac, and the size of QED corrections. 
The results are shown for the electron (left) and muon (right) channels.
}
\label{fig:Had_Wm_yW_Cor}
\end{figure}


\begin{figure}[!ht]
\setlength{\unitlength}{1mm}
\begin{picture}(160,87)
\put( -3,-1){\makebox(0,0)[lb]{
\epsfig{file=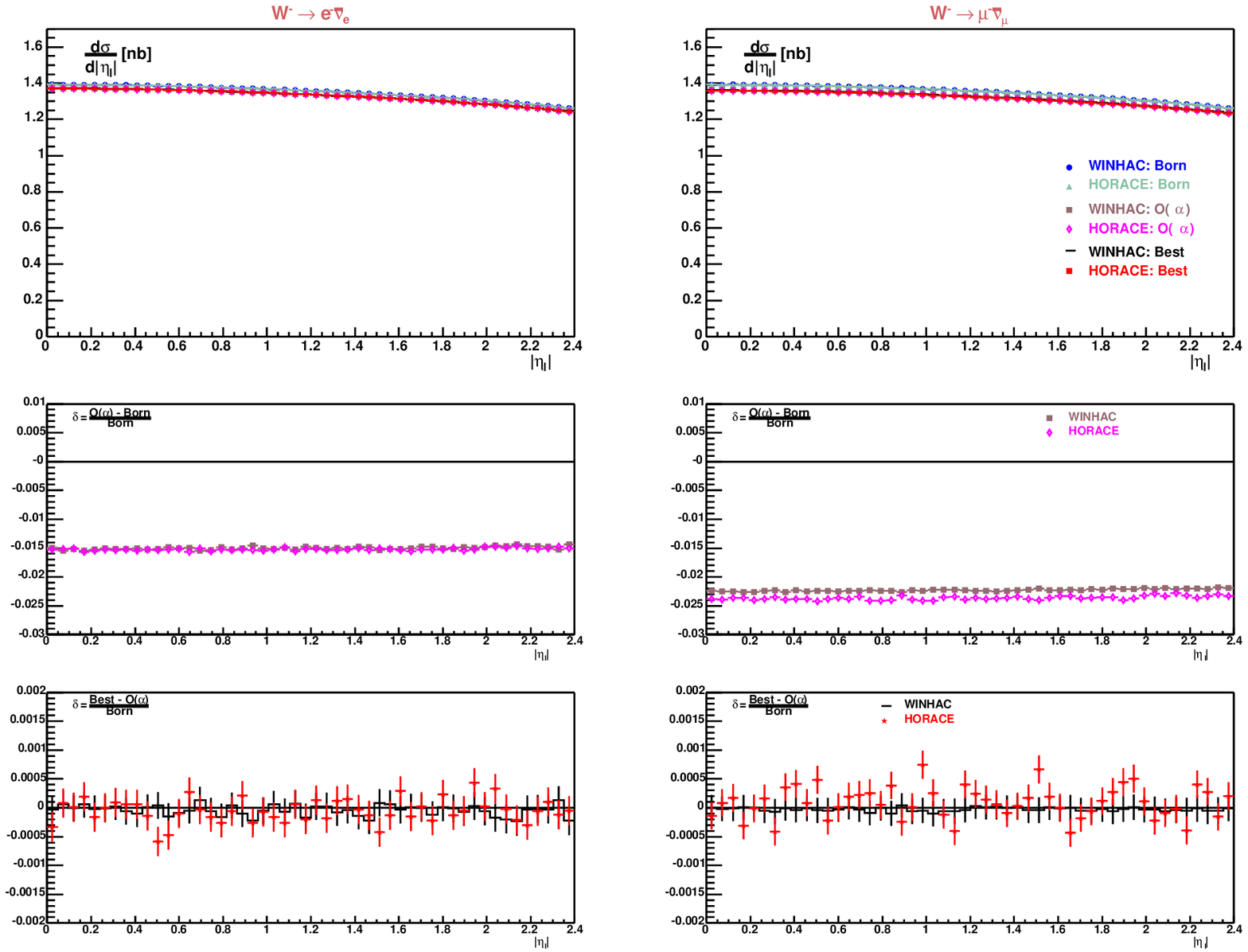,width=168mm,height=88mm}
}}
\end{picture}
\caption{\small\sf
Distributions of the charged lepton pseudorapidity at hadron level for Born, 
\oal\ and ``Best'' predictions of $W^-$ production 
from \horace\ and \winhac, and the size of QED corrections. 
The results are shown for the electron (left) and muon (right) channels.
}
\label{fig:Had_Wm_etal_Cor}
\end{figure}


\begin{figure}[!ht]
\setlength{\unitlength}{1mm}
\begin{picture}(160,87)
\put( -3,-1){\makebox(0,0)[lb]{
\epsfig{file=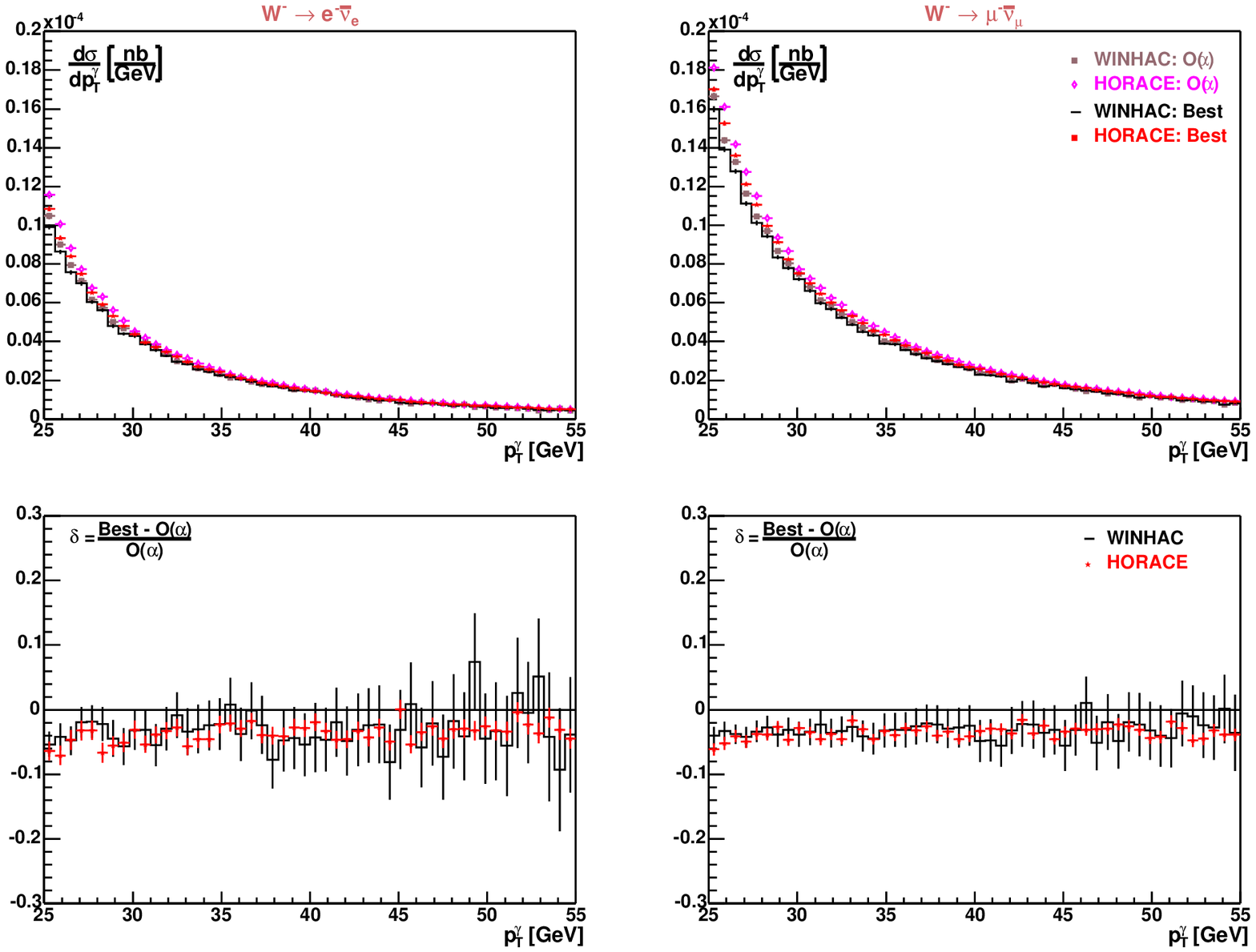,width=168mm,height=88mm}
}}
\end{picture}
\caption{\small\sf
Distributions of the hardest-photon transverse momentum at hadron level for  
\oal\ and ``Best'' predictions of $W^-$ production 
from \horace\ and \winhac, and the size of QED corrections. 
The results are shown for the electron (left) and muon (right) channels.
}
\label{fig:Had_Wm_pTg_Cor}
\end{figure}


\begin{figure}[!ht]
\setlength{\unitlength}{1mm}
\begin{picture}(160,87)
\put( -3,-1){\makebox(0,0)[lb]{
\epsfig{file=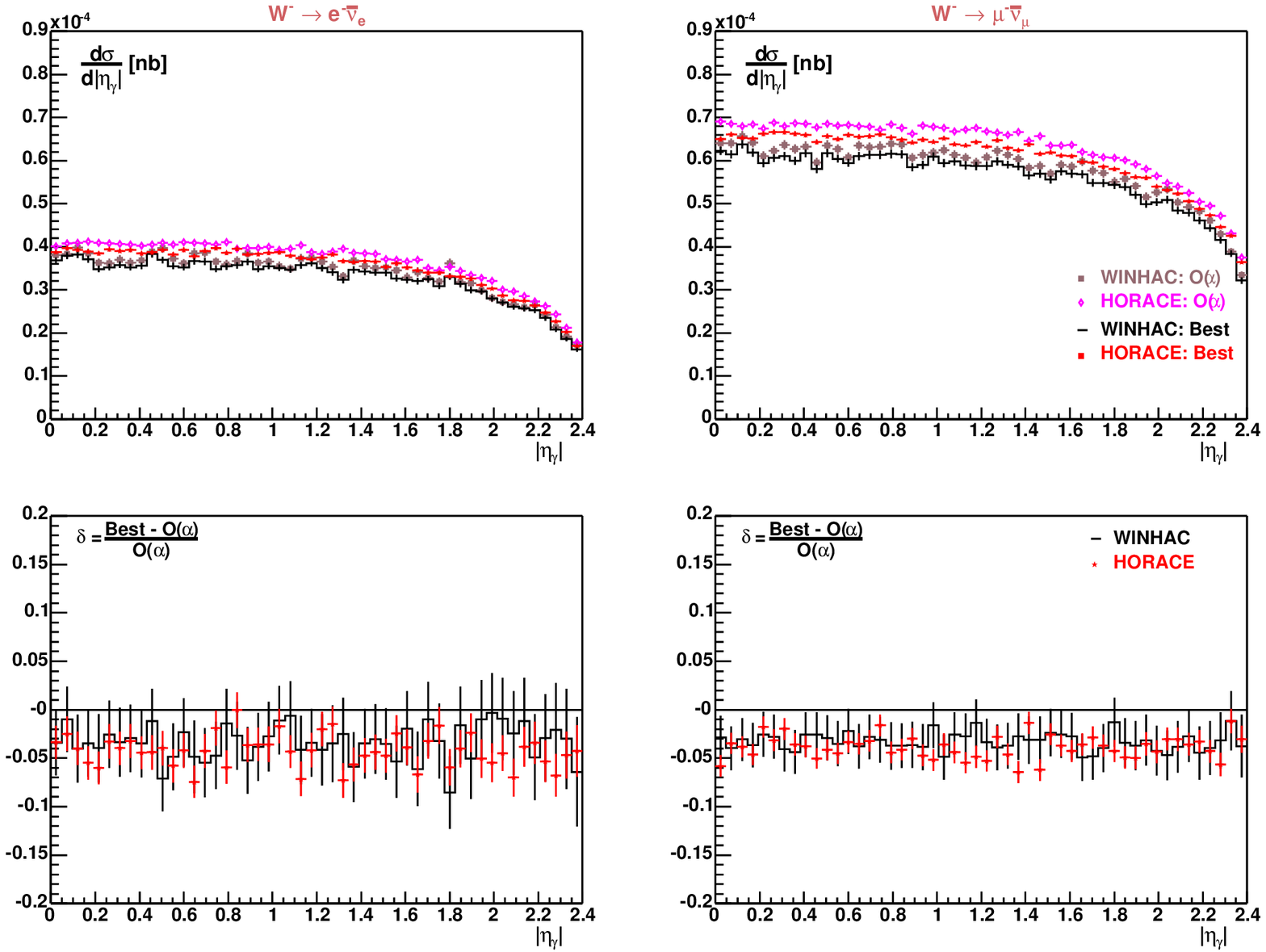,width=168mm,height=88mm}
}}
\end{picture}
\caption{\small\sf
Distributions of the hardest-photon pseudorapidity at hadron level
\oal\ and ``Best'' predictions of $W^-$ production 
from \horace\ and \winhac, and the size of QED corrections. 
The results are shown for the electron (left) and muon (right) channels.
}
\label{fig:Had_Wm_etag_Cor}
\end{figure}


%% file: hadxen.tex
%
\begin{figure}[!ht]
\setlength{\unitlength}{1mm}
\begin{picture}(160,83)
\put( -3,-2){\makebox(0,0)[lb]{
\epsfig{file=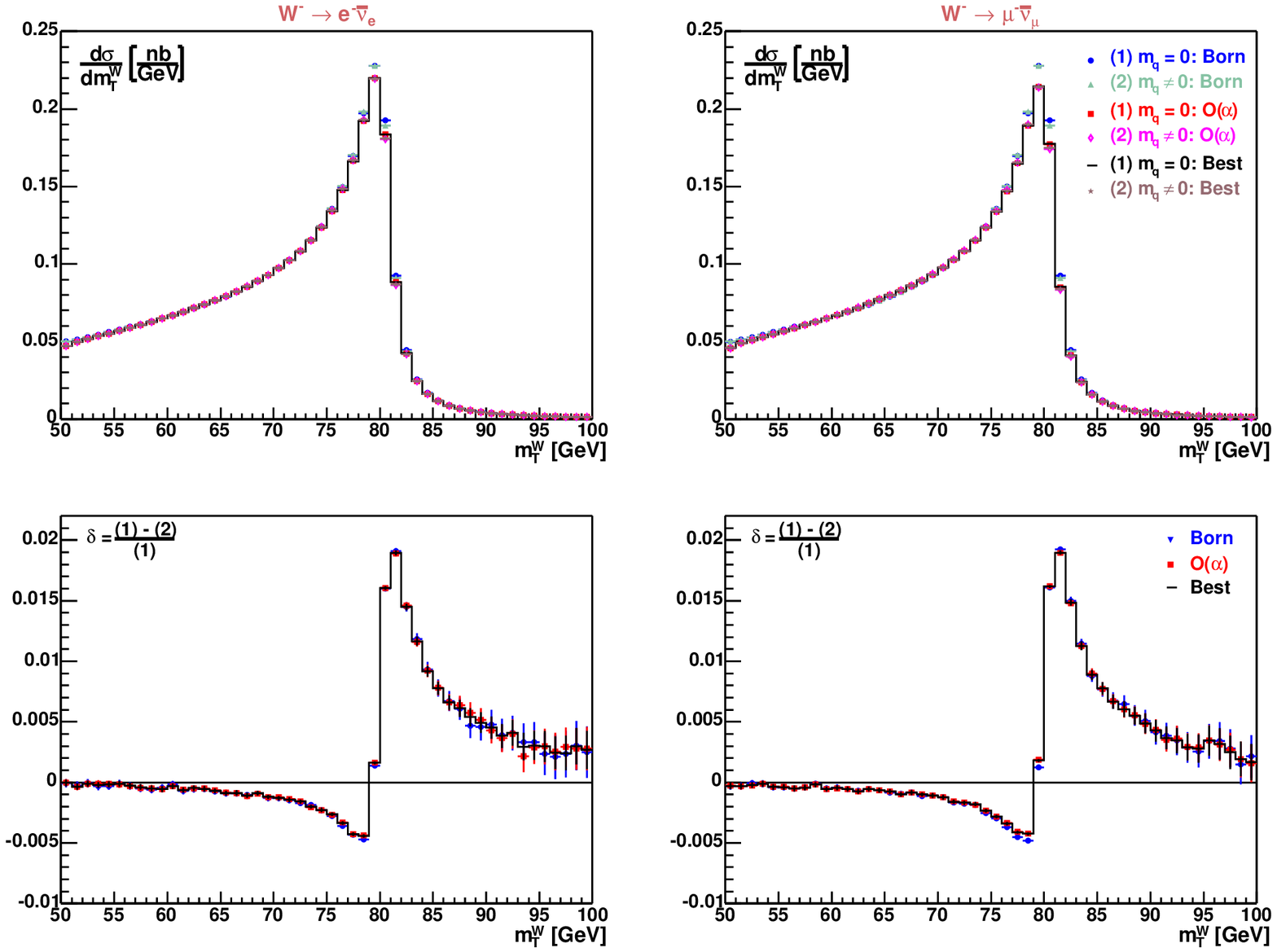,width=168mm,height=88mm}
}}
\end{picture}
\caption{\small\sf
The quark-mass effects in the ``energy-like scheme'' 
for the $W^-$ transverse mass distribution at the Born, \oal\ and ``Best'' 
levels, obtained from \winhac. 
The results are shown for the electron (left) and muon (right) channels.
}
\label{fig:xen_Wm_MTW}
\end{figure}
%
%
\begin{figure}[!ht]
\setlength{\unitlength}{1mm}
\begin{picture}(160,83)
\put( -3,-2){\makebox(0,0)[lb]{
\epsfig{file=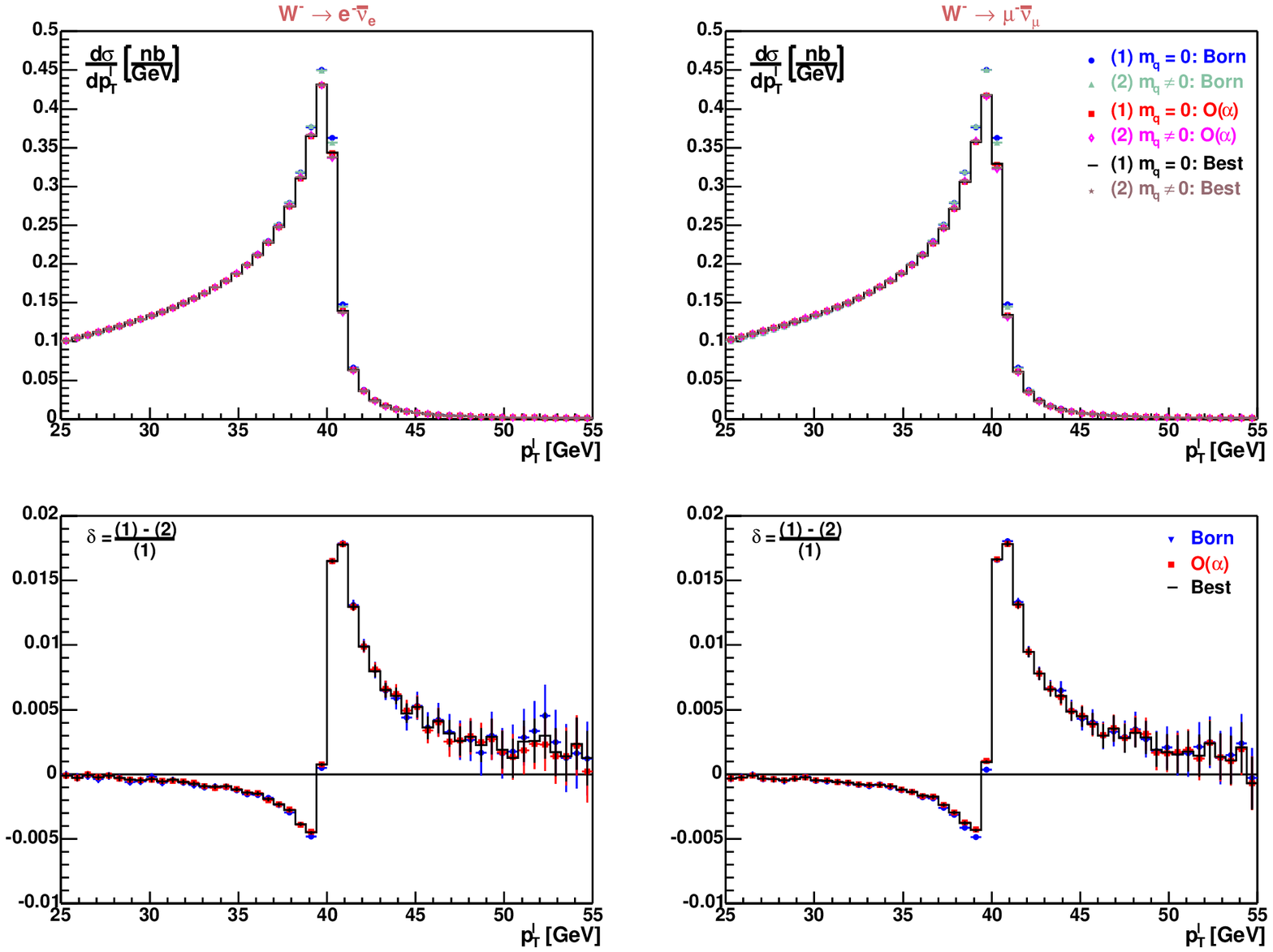,width=168mm,height=88mm}
}}
\end{picture}
\caption{\small\sf
The quark-mass effects in the ``energy-like scheme'' 
for the charged lepton transverse momentum distribution of $W^-$ production
at the Born, \oal\ and ``Best'' levels, obtained from \winhac. 
The results are shown for the electron (left) and muon (right) channels.
}
\label{fig:xen_Wm_pTl}
\end{figure}
%

\begin{figure}[!ht]
\setlength{\unitlength}{1mm}
\begin{picture}(160,88)
\put( -3,0){\makebox(0,0)[lb]{
\epsfig{file=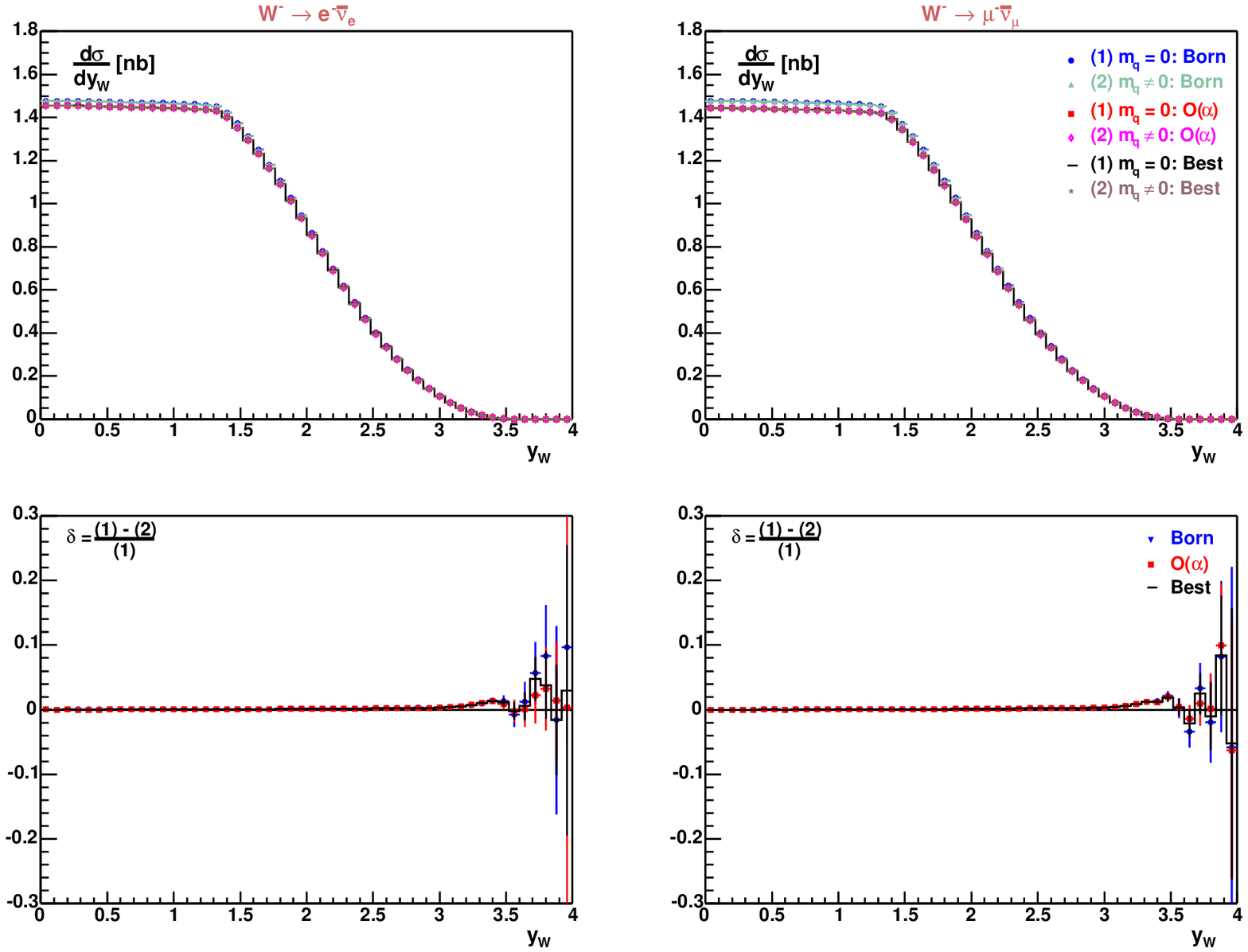,width=168mm,height=88mm}
}}
\end{picture}
\caption{\small\sf
The quark-mass effects in the ``energy-like scheme'' 
for the $W^-$ rapidity distribution at the Born, \oal\ and ``Best'' 
levels, obtained from \winhac. 
The results are shown for the electron (left) and muon (right) channels.
}
\label{fig:xen_Wm_yW}
\end{figure}
%
%
\begin{figure}[!ht]
\setlength{\unitlength}{1mm}
\begin{picture}(160,88)
\put( -3,0){\makebox(0,0)[lb]{
\epsfig{file=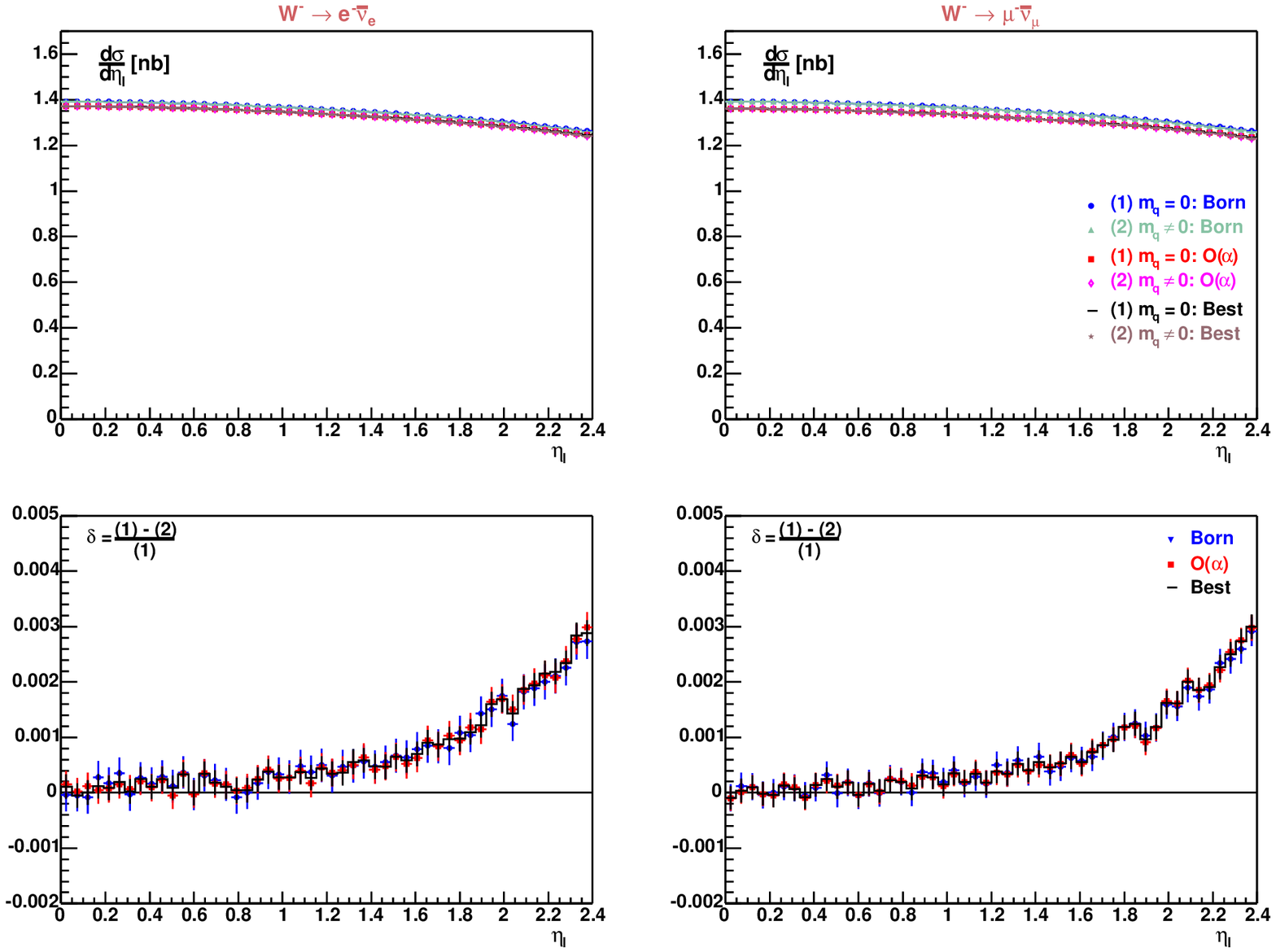,width=168mm,height=88mm}
}}
\end{picture}
\caption{\small\sf
The quark-mass effects in the ``energy-like scheme'' 
for the charged lepton  pseudorapidity distribution of $W^-$ production
at the Born, \oal\ and ``Best'' levels, obtained from \winhac. 
The results are shown for the electron (left) and muon (right) channels.
}
\label{fig:xen_Wm_etal}
\end{figure}

%% file: hadxpz.tex
\begin{figure}[!ht]
\setlength{\unitlength}{1mm}
\begin{picture}(160,83)
\put( -3,-2){\makebox(0,0)[lb]{
\epsfig{file=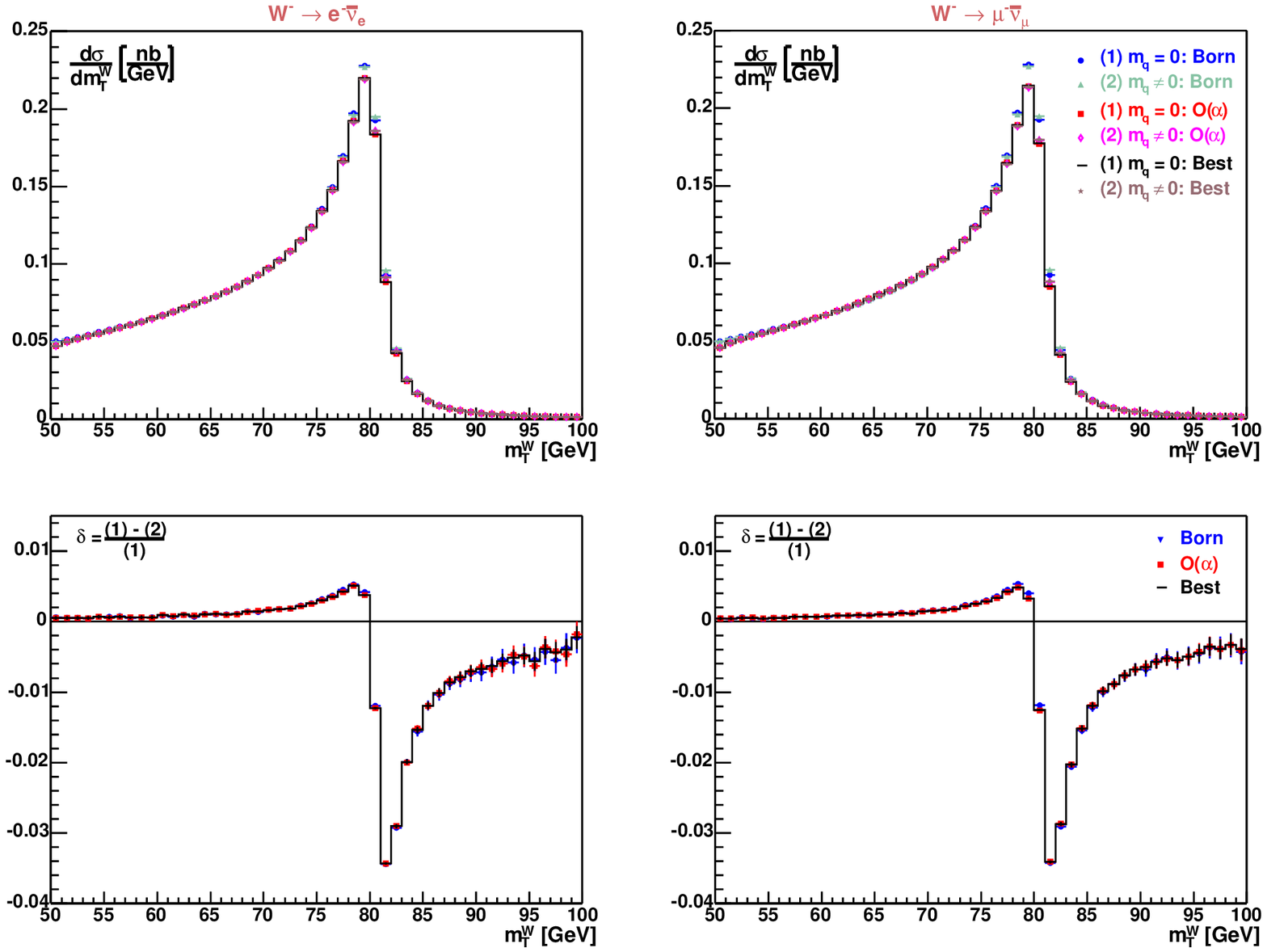,width=168mm,height=88mm}
}}
\end{picture}
\caption{\small\sf
The quark-mass effects in the ``momentum-like scheme'' 
for the $W^-$ transverse mass distribution at the Born, \oal\ and ``Best'' 
levels, obtained from \winhac. 
The results are shown for the electron (left) and muon (right) channels.
}
\label{fig:xpz_Wm_MTW}
\end{figure}
%
%
\begin{figure}[!ht]
\setlength{\unitlength}{1mm}
\begin{picture}(160,83)
\put( -3,-2){\makebox(0,0)[lb]{
\epsfig{file=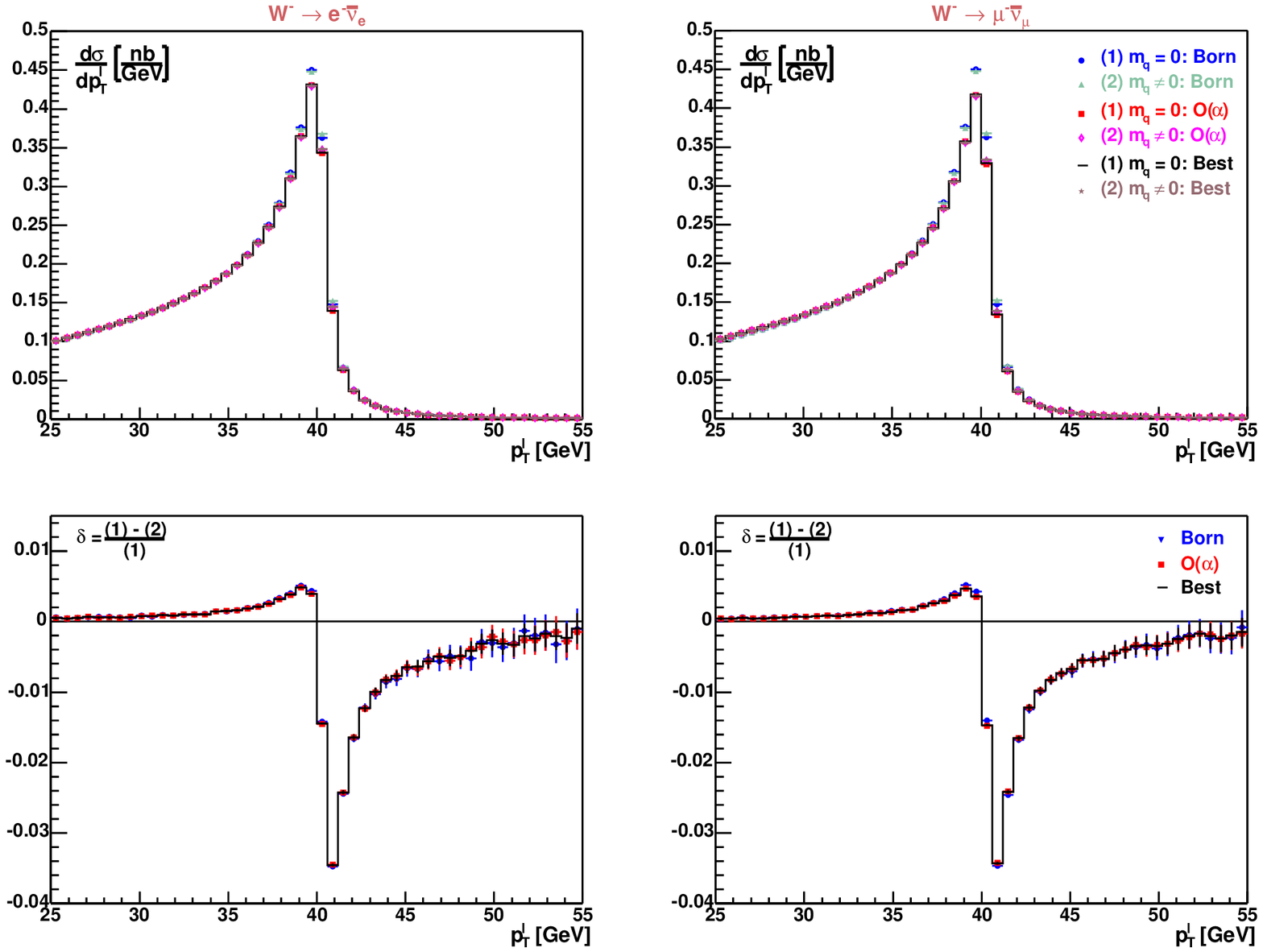,width=168mm,height=88mm}
}}
\end{picture}
\caption{\small\sf
The quark-mass effects in the ``momentum-like scheme'' 
for the charged lepton transverse momentum distribution of $W^-$ production
at the Born, \oal\ and ``Best'' levels, obtained from \winhac. 
The results are shown for the electron (left) and muon (right) channels.
}
\label{fig:xpz_Wm_pTl}
\end{figure}
%

\begin{figure}[!ht]
\setlength{\unitlength}{1mm}
\begin{picture}(160,88)
\put( -3,0){\makebox(0,0)[lb]{
\epsfig{file=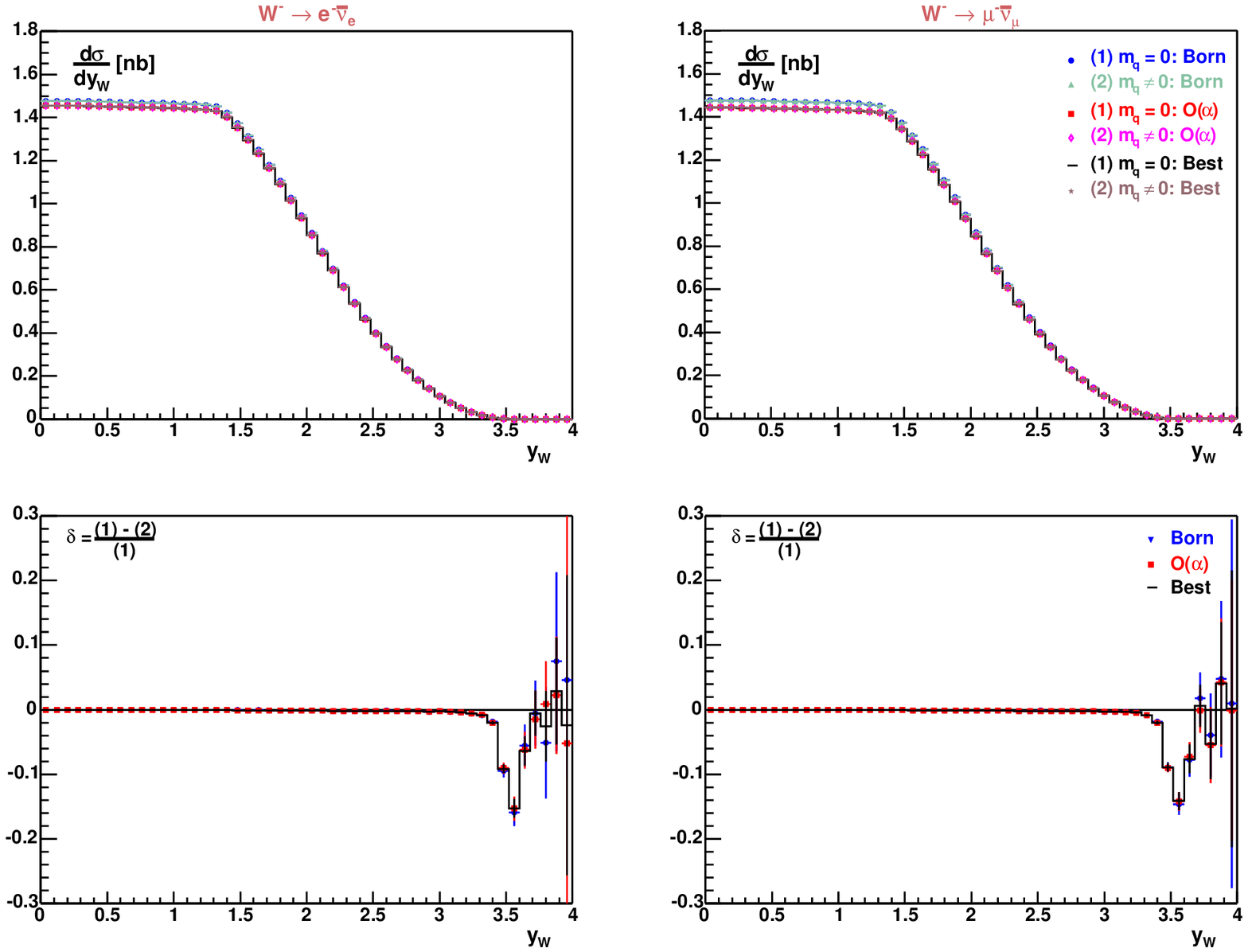,width=168mm,height=88mm}
}}
\end{picture}
\caption{\small\sf
The quark-mass effects in the ``momentum-like scheme'' 
for the $W^-$ rapidity distribution at the Born, \oal\ and ``Best'' 
levels, obtained from \winhac. 
The results are shown for the electron (left) and muon (right) channels.
}
\label{fig:xpz_Wm_yW}
\end{figure}
%
%
\begin{figure}[!ht]
\setlength{\unitlength}{1mm}
\begin{picture}(160,88)
\put( -3,0){\makebox(0,0)[lb]{
\epsfig{file=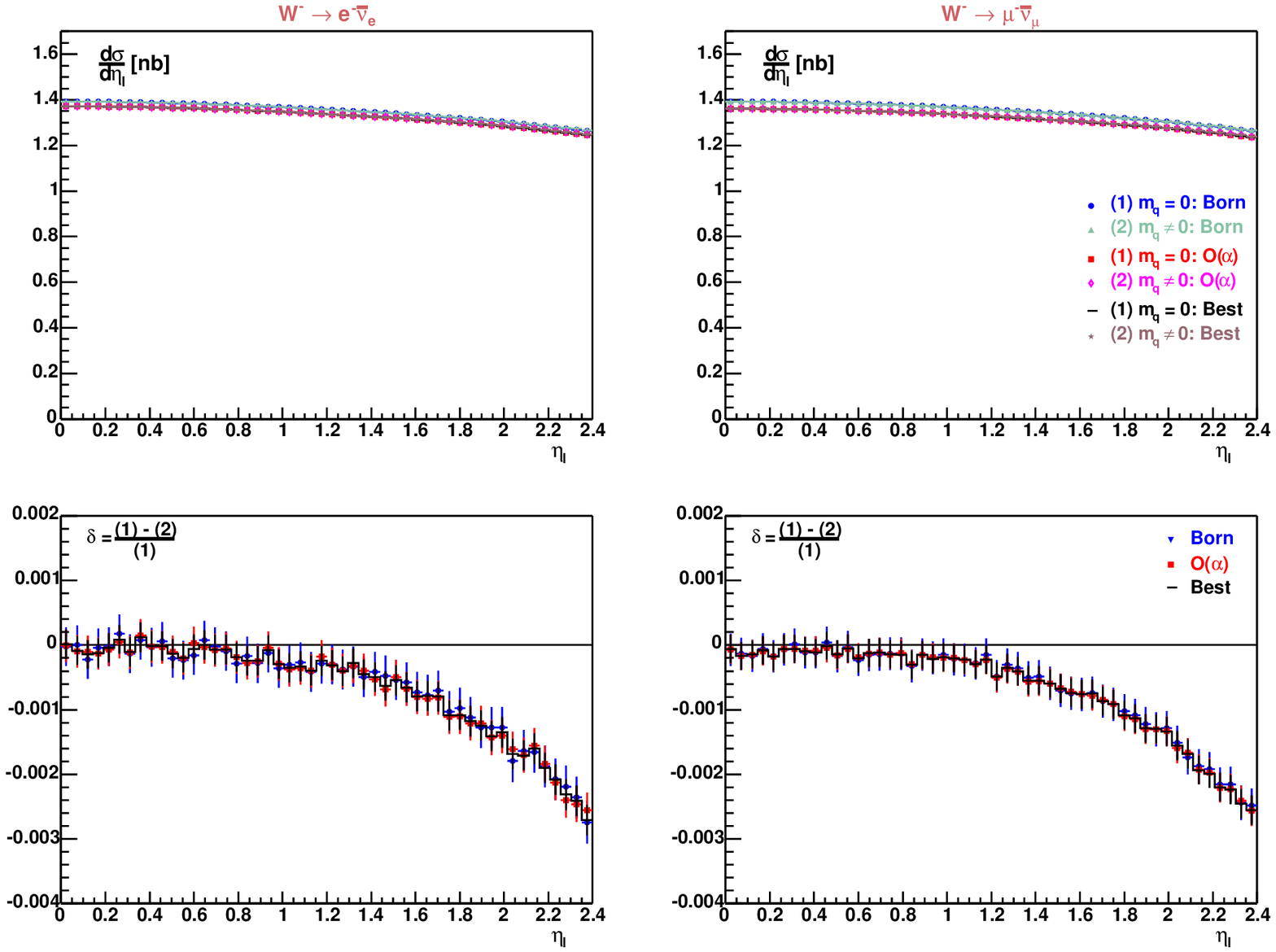,width=168mm,height=88mm}
}}
\end{picture}
\caption{\small\sf
The quark-mass effects in the ``momentum-like scheme'' 
for the charged lepton  pseudorapidity distribution of $W^-$ production
at the Born, \oal\ and ``Best'' levels, obtained from \winhac. 
The results are shown for the electron (left) and muon (right) channels.
}
\label{fig:xpz_Wm_etal}
\end{figure}

%% file: hadxlc.tex
\begin{figure}[!ht]
\setlength{\unitlength}{1mm}
\begin{picture}(160,83)
\put( -3,-2){\makebox(0,0)[lb]{
\epsfig{file=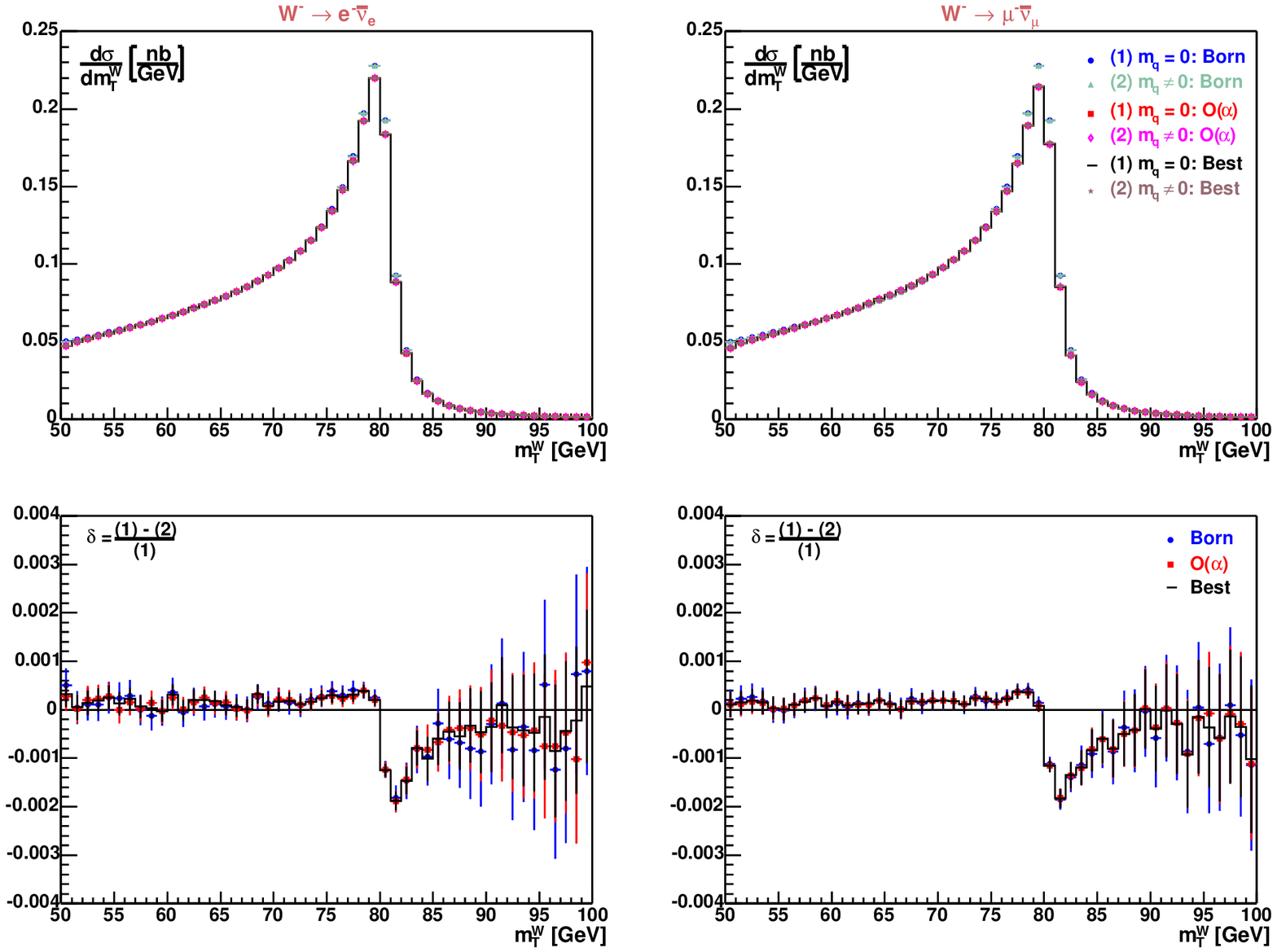,width=168mm,height=88mm}
}}
\end{picture}
\caption{\small\sf
The quark-mass effects in the ``light-cone-like scheme'' 
for the $W^-$ transverse mass distribution at the Born, \oal\ and ``Best'' 
levels, obtained from \winhac. 
The results are shown for the electron (left) and muon (right) channels.
}
\label{fig:xlc_Wm_MTW}
\end{figure}
%
%
\begin{figure}[!ht]
\setlength{\unitlength}{1mm}
\begin{picture}(160,83)
\put( -3,-2){\makebox(0,0)[lb]{
\epsfig{file=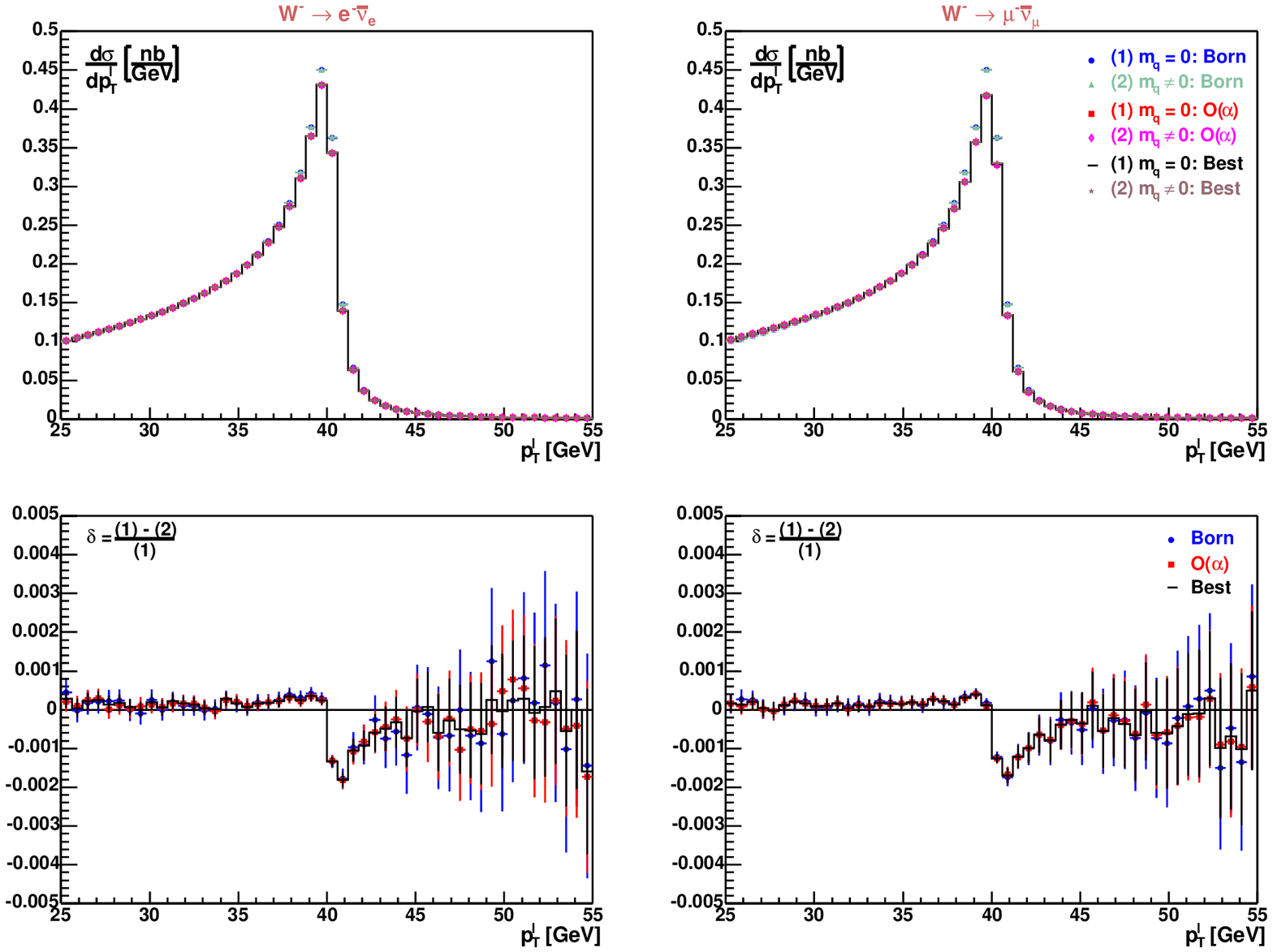,width=168mm,height=88mm}
}}
\end{picture}
\caption{\small\sf
The quark-mass effects in the ``light-cone-like scheme'' 
for the charged lepton transverse momentum distribution of $W^-$ production
at the Born, \oal\ and ``Best'' levels, obtained from \winhac. 
The results are shown for the electron (left) and muon (right) channels.
}
\label{fig:xlc_Wm_pTl}
\end{figure}
%

\begin{figure}[!ht]
\setlength{\unitlength}{1mm}
\begin{picture}(160,88)
\put( -3,0){\makebox(0,0)[lb]{
\epsfig{file=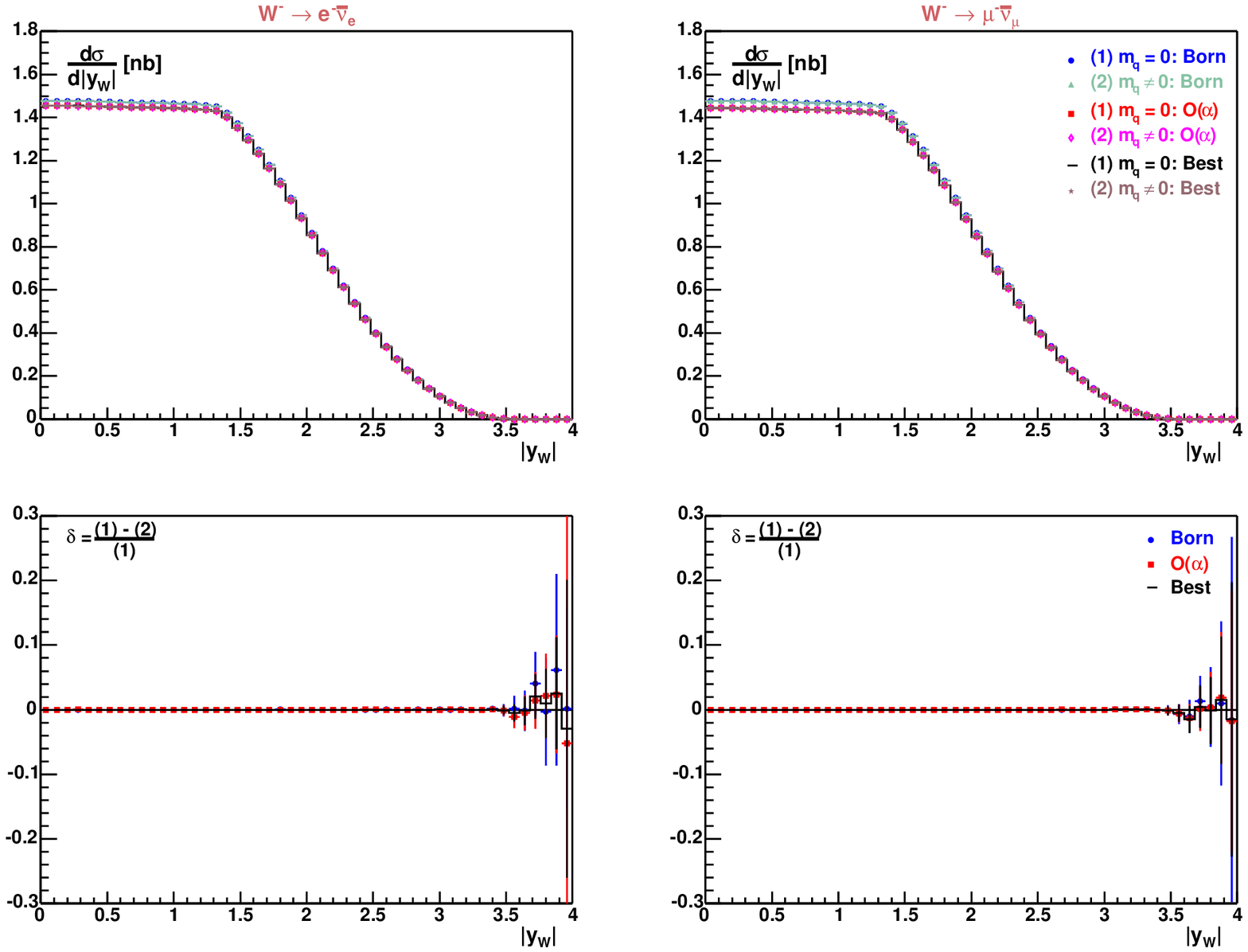,width=168mm,height=88mm}
}}
\end{picture}
\caption{\small\sf
The quark-mass effects in the ``light-cone-like scheme'' 
for the $W^-$ rapidity distribution at the Born, \oal\ and ``Best'' 
levels, obtained from \winhac. 
The results are shown for the electron (left) and muon (right) channels.
}
\label{fig:xlc_Wm_yW}
\end{figure}
%
%
\begin{figure}[!ht]
\setlength{\unitlength}{1mm}
\begin{picture}(160,88)
\put( -3,0){\makebox(0,0)[lb]{
\epsfig{file=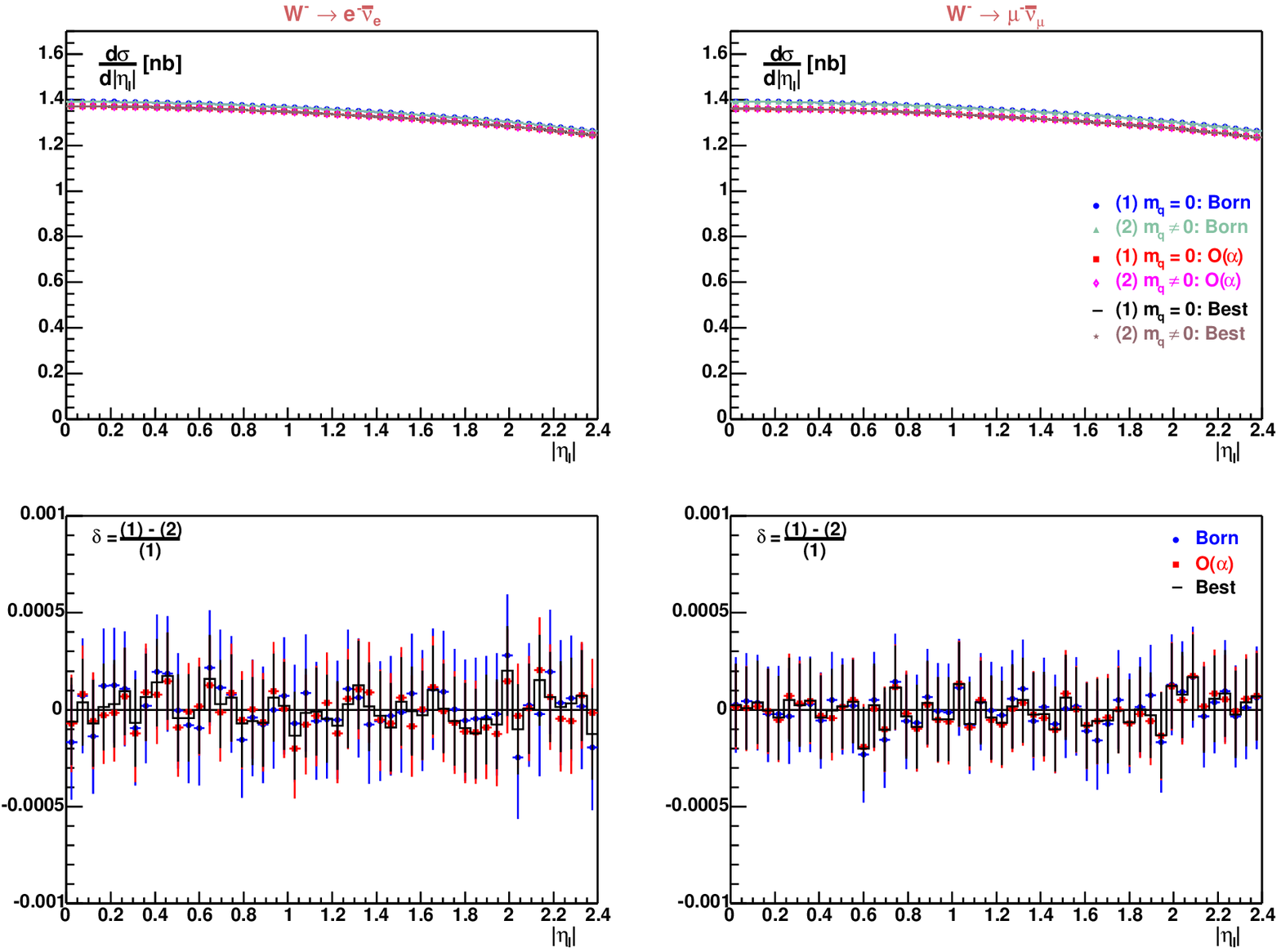,width=168mm,height=88mm}
}}
\end{picture}
\caption{\small\sf
The quark-mass effects in the ``light-cone-like scheme'' 
for the charged lepton  pseudorapidity distribution of $W^-$ production
at the Born, \oal\ and ``Best'' levels, obtained from \winhac. 
The results are shown for the electron (left) and muon (right) channels.
}
\label{fig:xlc_Wm_etal}
\end{figure}
%

%